\documentclass[aps,floats]{revtex4}
\usepackage{amsmath,amssymb}
\usepackage{graphicx,epsfig}

\begin{document}
\bibliographystyle {plain}

\def\oppropto{\mathop{\propto}} 
\def\opsimeq{\mathop{\simeq}}
\def\opoverderline{\mathop{\overline}}
\def\operarrow{\mathop{\longrightarrow}}
\def\opsim{\mathop{\sim}}

\def\fig#1#2{\includegraphics[height=#1]{#2}}
\def\figx#1#2{\includegraphics[width=#1]{#2}}


\title{ Equilibrium of disordered systems : 
constructing the appropriate valleys \\
in each sample via strong disorder renormalization in configuration space} 


 \author{ C\'ecile Monthus and Thomas Garel }
  \affiliation{Institut de Physique Th\'{e}orique, CNRS and 
CEA Saclay, 91191 Gif-sur-Yvette cedex, France}

\begin{abstract}
To describe the equilibrium properties of disordered systems and the possible emergence of various 'phases' at low temperature, we adopt here the 'broken ergodicity' point of view advocated in particular by Palmer [Adv. Phys. 31, 669 (1982)] : the aim is then to construct the valleys of configurations that become separated by diverging barriers and to study their relative weights, as well as their internal properties. To characterize the slow non-equilibrium dynamics of disordered systems, we have recently introduced in [C. Monthus and T. Garel, J. Phys. A 41, 255002 (2008) and arxiv:0804.1847] a strong disorder renormalization procedure in configuration space, based on the iterative elimination of the smallest barrier remaining in the system. In the present paper, we show how this renormalization procedure allows to construct the longest-lived valleys in each disordered sample, and to obtain their free-energies, energies and entropies. This explicit RG formulation is very general since it can be defined for any master equation, and it gives new insights into the main ingredients of the droplet scaling picture. As an application, we have followed numerically the RG flow for the case of a directed polymer in a two-dimensional random medium to obtain histograms of the free-energy, entropy and energy differences between the two longest-lived valleys in each sample.
\end{abstract}

\maketitle

\section{ Introduction }

In statistical physics, any large-scale universal behavior is expected to 
come from some underlying renormalization ('RG') procedure that eliminates
all the details of microscopic models. In the presence of quenched 
disorder, interesting universal scaling behaviors occur both at phase
transitions (as in pure systems) but also in the low-temperature 
disorder-dominated phases. Since the main property of frozen disorder
is to break the translational invariance, the most natural renormalization
procedures that allow to describe spatial heterogeneities
are a priori real-space RG procedures.
Among real-space renormalization procedures that have been introduced for
pure systems \cite{realspaceRG}, 
Migdal-Kadanoff block renormalizations \cite{MKRG} play a special role,
because they can be considered
 either as approximate renormalization procedures on hypercubic lattices,
or as exact renormalization procedures on certain hierarchical lattices
\cite{berker,hierarchical}. They have thus been used to study
various disordered models, such as the diluted Ising model \cite{diluted}, 
ferromagnetic random Potts model \cite{Kin_Dom,Der_Potts,andelman} and
spin-glasses \cite{young,mckay,Gardnersg,bray_moore,Ban_Bray,muriel,thill}.
For the case of spin-glasses,
the main output of these Migdal-Kadanoff RG studies
is that the probability distribution $P_L(J)$ of the effective
coupling $J$ on scale $L$ satisfies a closed RG equation
whose solution flows towards a fixed shape with a scale-dependent
width $L^{\theta}$ that defines the droplet exponent $\theta$.
To adapt this idea to hypercubic lattices, the notion of 'Domain-wall RG'
has been developed \cite{Banavar,McM,heidelberg,Fis_Hus}, where 
the droplet exponent $\theta$ is defined via the change of
the free-energy of a given sample when periodic boundary conditions 
are replaced by antiperiodic boundary conditions.
This definition can be used numerically in dimensions $d=2,3$ to obtain
a fixed shape for the probability distribution of this free-energy difference
when rescaled by $L^{\theta}$ \cite{Banavar,McM,heidelberg}.
Although this result is expected to come from some underlying RG procedure,
we are not aware of an explicit principle to define a
 real-space RG procedure for this distribution for hypercubic lattices
(apart from the Migdal-Kadanoff approximation that actually replaces
the hypercubic lattice by a hierarchical lattice).
Because of the difficulties
 to formulate an appropriate coarse-graining in real space
(see the discussion in section VI of ref. \cite{Fis_Hus}),
Fisher and Huse \cite{Fis_Hus} have preferred to formulate
their droplet scaling theory as 'a scaling Ansatz for the nature, statistics,
energetics and dynamics of the low-lying large scale excitations' [...]
'using RG ideas only to justify our scaling assumptions'.

The aim of the present paper is to show that the universal scaling
behaviors that appear in the low-temperature disorder-dominated phases
of disordered systems
can be understood within an explicit and consistent RG procedure 
{\it in configuration space}.
We adopt here the 'broken ergodicity' point of view \cite{Palmer,Palmer83}
(see more details in section \ref{bePalmer}), so that the aim
of the RG procedure is to construct the valleys of configurations
 that are separated by large barriers.
To study the non-equilibrium dynamics of disordered systems,
we have recently introduced a strong disorder RG procedure 
 to characterize the statistics of barriers on various scales and
 the extremely-slow logarithmic growth of the coherence
length $l(t) \sim (\ln t)^{1/\psi}$
\cite{rgshort,rglong}.
 In the present paper, we show how this method can be used to construct
 the valleys that are separated by barriers above a given threshold
and to study their statistical properties.

The paper is organized as follows.
In section \ref{coarsegrained}, we briefly recalled the 'broken ergodicity'
point of view and explain how a coarse-grained dynamics between valleys
allows to obtain their free-energy differences.
In section \ref{full}, we describe the explicit strong-disorder RG rules
for random master equations and discuss the properties of the valleys
that emerge. In particular, we explain how the free-energy differences between
valleys is constrained by a conservation law for ratios of renormalized
transition rates along the RG flow.
In section \ref{simplified}, we describe the 'simplified RG rules'
that are valid at large scales near
 'infinite disorder fixed points', and that correspond to the notion of
quasi-equilibrium inside each valley. 
Our conclusions are summarized in section \ref{conclusion}.

\section{ Defining valleys via some coarse-grained dynamics }

\label{coarsegrained}

\subsection{ Reminder on the 'ergodicity/broken ergodicity' point of view 
in statistical physics \cite{Palmer}}

\label{bePalmer}

The statistical physics of equilibrium is based 
on Boltzmann's ergodic principle, that states the equivalence
between the 'time average' of any observable $A$ over a sufficiently long time $t$
by an 'ensemble average' over microscopic
configurations $c$ of energies $e(c)$
\begin{eqnarray}
\frac{1}{t} \int_0^{t} d\tau A(\tau) \opsimeq_{t \to \infty}
 \sum_{c } A \left( c   \right) 
p_{eq} \left( c   \right)
\label{ergodic}
\end{eqnarray} 
where $p_{eq} \left( c   \right)$ represents
the Boltzmann measure at temperature $T=1/\beta$   
\begin{eqnarray}
p_{eq} \left( c   \right) = \frac{e^{- \beta e(c) } }
{Z_{tot} }
\label{boltzmann}
\end{eqnarray} 
and where $Z_{tot}$ is the partition function
\begin{eqnarray}
Z_{tot} \equiv \sum_c e^{- \beta e(c) }
\label{defzmicro}
\end{eqnarray}
Even if historically and physically, this dynamical interpretation
of the equilibrium is crucial, it is sometimes a bit 'forgotten'.
In particular to discuss the appearance of low temperature
symmetry broken phases in pure systems such as ferromagnets, 
it has become usual to reason only {\it statically} in terms of the properties
 of the Boltzmann measure in the thermodynamic limit,
but this way of thinking usually involves some 'cheating',
in the sense that one adds an infinitesimal magnetic external field,
or some special boundary conditions for the spins to select 
the possible 'pure states' that are obvious from the very beginning.
For disordered systems, many discussions
of the equilibrium are based on the same purely 'static' way of thinking,
but they face the very essential problem
that whenever disorder induces some frustration, 
the possible 'pure states' are not at all obvious 
because they depend on the realization of the randomness.
In particular for spin-glasses (see the reviews \cite{BY})
 which constitute the most studied frustrated disordered system, 
there exists a long-standing controversy on the nature and
the number of pure states,
as well as on the definition of appropriate order parameters.
The main descriptions are the mean-field-inspired 'replica-symmetry-breaking' 
point of view \cite{replica},
 the renormalization-inspired 'droplet' scaling picture
 \cite{McM,heidelberg,Fis_Hus},
 the more recent numerically-inspired 'Trival-Non Trivial'  \cite{tnt}
and 'state hierarchy of correlated spin domains' \cite{rammal,domany}
descriptions.
As a consequence, a purely static definition of pure states
based on the infinite-volume limit of finite-volume Boltzmann measures
which is simple for ferromagnets when the answer 
is known in advance, turns out to become very subtle
for frustrated disordered systems where the number
and the nature of pure states are not known in advance 
(see for instance the discussions in \cite{Fis_Hus_purestates,New_Stein}).
We thus feel that for such systems, a much clearer physical
description can be achieved by returning to the 
'historical' point of view of statistical mechanics
where the equilibrium is considered as the
stationary measure of some {\it dynamics},
so that the question on the number of 'phases' 
for the equilibrium becomes
a question of ergodicity-breaking for the dynamics :
are there valleys in configuration space
 that become separated by diverging barriers
in the thermodynamic limit and that keep nevertheless
finite free-energy differences ? 
This broken-ergodicity point of view is actually 
also the 'historical' point of view for the spin-glass problem :
in their original paper \cite{EA}, 
Edwards and Anderson have defined their order parameter by
the following sentence :
`` if on one observation a particular spin is $S_i(0)$,
then if it is studied again a long time later, there is a non-vanishing
probability that $S_i(t)$ will point in the same direction''.
The importance of this definition has been further emphasized 
by Anderson in \cite{And_houches} : ``If the spins are going to polarize
in a particular random function [...], we had better not try to characterize
the order by some kind of long-ranged order in space,
or by some kind of order parameter defined in space,
but  { \it we must approach it from a pure
non-ergodic point of view, as a long-range order in the time alone } :
 if the system has a certain order at $t=0$, then as $t \to + \infty$ 
there remain a finite memory of that order''
\begin{eqnarray}
q_i = \lim_{t \to \infty} < S_i(0) S_i(t) > 
\label{EA_time}
\end{eqnarray} 
Many works have then tried to characterize the phase space structure
of spin-glasses via dynamical studies, in particular by
measuring numerically the distance between two
configurations that are submitted to the same thermal noise
to detect the presence of different valleys that become
separated by diverging barriers in the thermodynamic limit 
\cite{Der_damagespreading}. Since the works on spin-glasses
 that are based on a 'dynamical' point of view of the equilibrium
are too numerous to be summarized here, we will stop here this reminder,
and we refer the reader to the papers of Palmer
where the issues related to 'broken ergodicity'
are discussed in detail, both for
statistical physics models in general \cite{Palmer}
and for spin-glasses in particular \cite{Palmer,Palmer83}.
Once one has adopted this 'broken ergodicity' point of view,
 the aim is to construct the valleys
that tend to confine the dynamics in the thermodynamic limit.
The starting point is thus some microscopic dynamics in phase space.

\subsection{ Microscopic dynamics with detailed balance}

In statistical physics, it is convenient to define the dynamics via a
 master equation describing the evolution of the
probability $p_t (c ) $ to be in a microscopic configuration $c$ at time t
\begin{eqnarray}
\frac{ dp_t \left(c \right) }{dt}
= \sum_{ c '} p_t \left(c' \right)
 w \left(c' \to  c  \right) 
 -  p_t \left(c \right) w_{out} \left( c \right)
\label{master}
\end{eqnarray}
The notation  
$ w \left(c' \to  c  \right) $ 
represents the transition rate per unit time from configuration 
$c'$ to $c$, and the notation
\begin{eqnarray}
w_{out} \left( c \right)  \equiv
 \sum_{ c '} w \left(c \to  c' \right) 
\label{wcout}
\end{eqnarray}
represents the total exit rate out of configuration $c$.
To ensure that any finite system will converge towards thermal equilibrium
in the limit of infinite time,
we consider as usual 
that the transition rates satisfy the detailed balance property
\begin{eqnarray}
 e^{- \beta e(c) } w \left( c \to c'  \right)
= e^{- \beta e(c') } w \left( c' \to c  \right)
\label{detailedbalance}
\end{eqnarray}
Then, the Boltzmann equilibrium distribution of Eq \ref{boltzmann}
is the stationary solution of the master equation of Eq. \ref{master}.

\subsection{ Notion of coarse-grained dynamics between valleys }

We now assume that the microscopic master equation of Eq. \ref{master}
can be coarse-grained into a renormalized master equation
\begin{eqnarray}
\frac{ dP_t \left( C \right) }{dt}
= \sum_{ C '} P_t \left( C' \right)
 W \left( C' \to  C \right) 
 -  P_t \left( C \right) W_{out} \left( C \right)
\label{masterR}
\end{eqnarray}
between 'renormalized configurations' denoted by $C$ 
(not to be confused with the microscopic configurations denoted by $c$)
 in terms of renormalized transition rates $W$ 
(not to be confused with the microscopic transition rates denoted by $w$).
An explicit procedure to renormalize master equation
will be discussed in details in the next section \ref{full},
but here in the remaining of this section,
our aim is to discuss the general meaning 
of any such coarse-grained dynamics.
From now on, 'renormalized configurations' will be called 'valleys'
to simplify the formulation.

\subsection{ Definition of the free-energies $F(C)$ of valleys }

The stationary solution $P_{st}(C)$ of the renormalized master equation
of Eq. \ref{masterR} is then fixed (up to a normalization constant)
by the ratios of the renormalized transition rates
\begin{eqnarray}
\frac{ P_{st}(C) }{P_{st}(C')}
= \frac{W \left( C' \to  C \right)}{W \left( C \to  C' \right)}
\label{pstatioR}
\end{eqnarray}

In the statistical physics of equilibrium, it has been understood
since Einstein that the probability of some fluctuation
is determined by the free-energy cost $\Delta F(Fluct)$ of this fluctuation 
\begin{eqnarray}
{\rm Prob}_{eq} (Fluct) \propto e^{- \beta \Delta F(Fluct)} 
\label{fluct}
\end{eqnarray}
This property is nowadays considered as sufficiently 'fundamental'
to be used to extend the notion of free-energy to non-equilibrium situations :
 the 'out-of-equilibrium' free-energy is then defined from
 the large deviation function that govern the probability of fluctuations
(see the recent review \cite{derrida_noneq} and references therein
for more detailed explanations and examples).

For our present problem in any finite system,
 the coarse-grained master equation will converge
towards the stationary distribution $P_{st}(C)$ in the limit of infinite-time.
 It is thus natural to define the free-energy $F(C)$
of the valley $C$ from the 
stationary solution $P_{st}(C)$ of the renormalized master equation by
\begin{eqnarray}
 P_{st}(C) = \frac{ e^{- \beta F(C)} }{ Z_R}
\label{freeenergyV}
\end{eqnarray}
with the normalization $Z_R$ ensuring that $\sum_C  P_{st}(C) =1$.
In particular, the ratios of the renormalized 
transition rates of Eq. \ref{pstatioR}
determines directly the free-energy differences between valleys 
\begin{eqnarray}
 e^{- \beta (F(C)-F(C'))} 
= \frac{W \left( C' \to  C \right)}{W \left( C \to  C' \right)}
\label{freediffWR}
\end{eqnarray}

It is very important to stress here that the free-energies of the valleys
have been defined in Eqs \ref{freeenergyV} and \ref{freediffWR}
from the coarse-grained dynamics between valleys,
i.e. from the {\it inter-valleys } dynamics.
But we have made no statement yet about what happens inside each valley,
because the intra-valley properties actually depend on the 
explicit principle that is used to renormalize the master equation.
We will thus rediscuss this point in the next sections 
for the explicit strong-disorder renormalization procedure
that we will use to renormalize the master equation.

\subsection{ Inter-valleys entropy $S^{inter}$ and 
  valleys weights statistics }

Once the stationary distribution $P_{st}(C)$
of the renormalized master equation is know, it is natural to
introduce the following inter-valleys entropy $S^{inter}$  
\begin{eqnarray}
 S^{inter}= - \sum_{C} P_{st}(C) \ln  P_{st}(C) 
\label{sinter}
\end{eqnarray}
This quantity is called 'complexity' in \cite{Palmer,Palmer83},
but since this word is used with different meanings,
we will keep the name 'inter-valley entropy' for clarity.

To better characterize the statistics of the valleys weights,
it is interesting to introduce the generalized moments \cite{yk}
\begin{eqnarray}
 Y_k=  \sum_{C} \left[ P_{st}(C) \right]^k
\label{defyk}
\end{eqnarray}
(one has the normalization $Y_{k=1}=1$, and
the entropy $S^{inter}$ of Eq. \ref{sinter} corresponds to 
$S^{inter} =- \partial Y_k \vert_{k=1}$).
Results on the valley weights statistics in mean field spin-glasses
(in the Random Energy Model and in the SK model)
and in other statistical physics models are reviewed in \cite{yk}.
More detailed studies on probability distributions of
the $Y_k$ can be found in  \cite{Der_Fly}.

\section{ Strong disorder renormalization of master equations}

\label{full}

In the present section, we explained 
why the strong disordered renormalization of master equation introduced
in \cite{rgshort} is the appropriate way to construct the coarse-grained
dynamics between valleys starting from the microscopic master equation.
Strong disorder renormalization 
(see \cite{review} for a review) is a very specific type of RG
that has been first developed in the field of quantum spins :
the RG rules of Ma and Dasgupta \cite{madasgupta} 
have been put on a firm ground by D.S. Fisher 
who introduced the crucial idea of ``infinite disorder'' fixed point
where the method becomes asymptotically exact,
and who computed explicitly exact 
critical exponents and scaling functions 
for one-dimensional disordered quantum spin chains \cite{dsf}.
This method has thus generated a lot of activity for various
disordered quantum models \cite{review}. It has been then
successfully applied to
various classical disordered dynamical models,
such as random walks in random media \cite{sinairg,sinaibiasdirectedtraprg},
reaction-diffusion in a random medium \cite{readiffrg}, 
coarsening dynamics of classical spin chains \cite{rfimrg}, 
trap models \cite{traprg}, random vibrational networks \cite{vibrational},
absorbing state phase transitions \cite{contactrg},
zero range processes \cite{zerorangerg} and 
exclusion processes  \cite{exclusionrg}.
In all these cases, the strong disorder RG rules 
have been formulated {\it in real space},
with specific rules depending on the problem.
For more complex systems where
 the formulation of strong disorder RG rules
has not been possible in real space, 
we have recently proposed in \cite{rgshort} a strong disorder 
RG procedure { \it in configuration space} that can be
 defined for any master equation. In the remaining of this section,
we first recall this procedure, and then discuss its consequence 
for the free-energy differences that can be extracted 
from the renormalized transition rates (see Eq. \ref{freediffWR}),
and for the energies of the valleys.

\subsection{ Reminder on the 'full' strong disorder RG rules \cite{rgshort} }

For dynamical models, the aim of any renormalization procedure
is to integrate over 'fast' processes to obtain effective properties 
of 'slow' processes.
 The general idea of 'strong renormalization' for dynamical models
consists in eliminating iteratively the 'fastest' process.
The RG procedure introduced
in \cite{rgshort} can be summarized as follows
( see \cite{rgshort,rglong} for more explanations ) :

(1) find the configuration ${ C}^*$ with the largest exit rate $W^*_{out}$
\begin{eqnarray}
W^*_{out} = W_{out} \left( { C}^* \right)
 \equiv  {\rm max}_{{ C}} \left[  W_{out} \left( { C} \right) \right]
\label{defwmax}
\end{eqnarray}

(2) find the neighbors $({ C}_1,{ C}_2,...,{ C}_n)$
 of configuration ${ C}^*$, 
i.e. the configurations that were related 
via positive rates $W({ C}^* \to { C}_i )>0$ and
 $W({ C}_i\to { C}^*)>0$
to the decimated configuration ${ C}^*$
(here we will assume for the simplicity of the discussion,
 and because it is usually the case in
statistical physics models, that if a transition has 
a strictly positive rate, the reverse transition has also
a strictly positive rate).
For each neighbor configuration ${ C}_i$ with
$i \in (1,..,n)$, update the transition rate to go to
the configuration ${ C}_j$ with $j \in (1,..,n)$ and $j \neq i$
 according to
\begin{eqnarray}
W^{new}({ C}_i \to { C}_j )=W({ C}_i \to { C}_j )
+ W({ C}_i \to { C}^* ) \times  \pi_{{ C}^*} \left({ C}_j \right) 
\label{wijnew}
\end{eqnarray}
where the first term represents the 'old' transition rate (possibly zero),
and the second term represents the transition 
via the decimated configuration ${ C}^*$ :
the factor $W({ C}_i \to { C}^* ) $ takes into account 
the transition rate to ${ C}^*$ and the term
\begin{eqnarray}
\pi_{{ C}^*} \left({ C}_j \right) = 
\frac{W \left({ C}^*  \to { C}_j \right)}{W_{out} \left( { C}^*\right)}
\label{picstar}
\end{eqnarray}
represents the probability
 to make a transition towards ${ C}_j$
when in ${ C}^*$.
The $2 n$ rates $W({ C}^* \to { C}_i )$
 and $W({ C}_i \to { C}^*)$ then
 disappear with the decimated configuration ${ C}^*$.
Note that the rule of Eq. \ref{wijnew} 
has been recently proposed in \cite{Pigo}
to eliminate 'fast states'  from various dynamical problems 
with two very separated time scales.
The physical interpretation of this rule is as follows :
the time spent in the decimated configuration ${ C}^*$ is neglected
with respect to the other time scales remaining in the system.
The validity of this approximation within the present framework
will be discussed in detail below in \ref{strongdisorderfixedpoint}.
The interesting equivalence of the rule of Eq \ref{picstar}
 with the well-known 'adiabatic' approximation \cite{Pigo}
will be recalled in section \ref{adiabatic} to 
clarify what is really 
assumed physically for the decimated renormalized configuration $C^*$.

(3) update the exit rates out of the neighbors ${ C}_i$, with $i=1,..,n$
either with
the definition
\begin{eqnarray}
W^{new}_{out}({ C}_i)   = 
\sum_{ C} W^{new}({ C}_i \to { C} )
\label{wioutnewactualisation}
\end{eqnarray}
or with the equivalent rule that can be deduced from Eq. \ref{wijnew}
 \begin{eqnarray}
W^{new}_{out}({ C}_i) = W_{out}({ C}_i) 
 - W({ C}_i \to { C}^* ) 
\frac{ W({ C}^* \to { C}_i )}
{W^{*}_{out}}
\label{wioutnew}
\end{eqnarray}
The physical meaning of this rule is the following.
The exit rate out of the configuration ${ C}_i$ decays because 
the previous transition towards ${ C}^*$ can lead to an immediate return
towards ${ C}_i$ with probability 
$\pi_{{ C}^*} \left({ C}_i\right) =
\frac{ W({ C}^* \to { C}_i )}
{W^{*}_{out}} $. After the decimation of the configuration ${ C}^*$,
this process is not considered as an 'exit' process anymore, but as a
residence process in the configuration ${ C}_i$.
This point is very important to understand the
 meaning of the renormalization procedure :
the remaining configurations at a given renormalization scale are 
'formally' microscopic configurations 
of the initial master equation (Eq. \ref{master}),
but each of these remaining microscopic configuration
 actually represents some 'valley' in configuration space
that takes into account all the previously decimated configurations.

(4) return to point (1).

Note that in practice, the renormalized rates $W({ C} \to { C}' )$
can rapidly  become very small as a consequence of the multiplicative structure
of the renormalization rule of Eq \ref{wijnew}. This means that the appropriate
 variables are the logarithms of the transition rates, that we will call 'barriers' in the remaining
of this paper. The barrier $B ({ C} \to { C}' )$ from ${ C}$  to ${ C}' $ is defined by
\begin{eqnarray}
B ({ C} \to { C}' )\equiv - \ln W({ C} \to { C}' )
\label{defb}
\end{eqnarray}
and similarly the exit barrier out of configuration ${ C}$ is defined by
\begin{eqnarray}
B_{out} ({ C} )\equiv - \ln W_{out}({ C} )
\label{defbout}
\end{eqnarray}
Note that a very important advantage of this formulation in terms
of the renormalized transition rates of the master equation is that 
the renormalized barriers take into account the true 'barriers'
of the dynamics, whatever their origin which can be
 either energetic or entropic.

\subsection{ Notion of 'infinite disorder fixed point' 
and asymptotic exactness of the RG rules  }

\label{strongdisorderfixedpoint}

As mentioned above, the approximation made
 in the renormalization rule of Eq. \ref{wijnew}
consists in neglecting the time spent 
in the decimated configuration ${ C}^*$ 
with respect to the other time scales remaining in the system. 
In the present framework, this means that the maximal exit rate chosen
 in Eq \ref{defwmax}
should be well separated from the exit rates of the neighboring
 configurations ${ C}_i$.
The crucial idea of 'infinite disorder fixed point' \cite{dsf,review}
is that even if this approximation is not perfect during the
 first steps of the renormalization,
this approximation will become better and better 
 at large time scale if the probability distribution
 of the remaining exit rates
becomes broader and broader upon iteration. 
More precisely,  if the renormalization scale $\Gamma$ 
is defined as the exit barrier of the
 last decimated configuration ${ C^*}$
\begin{eqnarray}
\Gamma= B_{out} ({ C^*} )\equiv - \ln W_{out}^*
\label{defgamma}
\end{eqnarray}
one expects that the probability distribution
 of the remaining exit barrier $B_{out} \geq \Gamma$ will 
converge towards some scaling form
\begin{eqnarray}
P_{\Gamma} ( B_{out} -\Gamma ) \opsimeq_{ \Gamma \to \infty} 
  \frac{1}{\sigma(\Gamma) } {\hat P} 
\left( \frac{B_{out} - \Gamma}{\sigma(\Gamma) } \right)
\label{pgammabout}
\end{eqnarray}
where ${\hat P} $ is the fixed point probability distribution, 
and where $\sigma(\Gamma)$ is the appropriate scaling factor
for the width.
The notion of 'infinite disorder fixed point' 
means that the width $\sigma(\Gamma)$ grows indefinitely with the
renormalization scale $\Gamma$
\begin{eqnarray}
\sigma(\Gamma) \opsimeq_{ \Gamma \to \infty} +\infty
\label{infinite}
\end{eqnarray}
Whenever this 'infinite disorder fixed point' condition 
 is satisfied, the strong disorder renormalization
procedure becomes asymptotically exact at large scales.
In previously known cases of infinite disorder fixed
 points where calculations can be done explicitly
\cite{review},  the scale $\sigma(\Gamma)$ has been 
found to grow linearly $
\sigma(\Gamma) \opsimeq_{ \Gamma \to \infty} \Gamma $.
This behavior means that the cut-off $\Gamma$ is the
 only characteristic scale and thus describes some critical 
point \cite{review}.
For the present procedure concerning the dynamics in disordered models,
this property means some 'criticality in the time direction',
i.e. the absence of any characteristic time scale between the microscopic scale
and the macroscopic equilibrium time of the full disordered sample
(see \cite{rglong} for more detailed discussions).

For the present strong disorder renormalization of a master equation,
the convergence towards an 'infinite disorder fixed point'  
will depend on the initial condition of the transition rates,
i.e. on the model (and on the temperature if there are phase transitions).
However, the form of the RG rules of Eq \ref{wijnew}
 is sufficiently similar to the usual Ma-Dasgupta rules
 \cite{review} to think that 
 the convergence towards some infinite disorder fixed point
 should be realized in
 a very broad class of disordered systems in their glassy phase.
In practice, it should be checked numerically for each model of interest.
We refer to \cite{rgshort,rglong} where this scenario has been checked
numerically for the dynamics
 of a directed polymer in a two dimensional random medium.

\subsection{ RG flow for the free-energy differences between valleys }

We have explained above how the free-energy difference between 
two valleys can be obtained from the ratio of the two
transition rates between them (Eq. \ref{freediffWR}).
It is thus convenient to introduce the following
notation for these ratios
\begin{eqnarray}
R \left({ C}_i  \to { C}_j \right)\equiv \frac{W \left({ C}_i  \to { C}_j \right)}
{W \left({ C}_j  \to { C}_i \right)}
\label{ratioij}
\end{eqnarray}
 to see how they evolve upon renormalization.
In the initial condition (RG scale $\Gamma=0$ ) corresponding to the
microscopic master equation of Eq. \ref{master},
the values of all these ratios are fixed by the detailed balance condition
of Eq \ref{detailedbalance} 
\begin{eqnarray}
R^{\Gamma=0}\left(c_i  \to c_j \right) \equiv \frac{w \left(c_i  \to c_j \right)}
{w \left(c_j  \to c_i \right)} = e^{- \beta (e(c_j)-e(c_i))}
\label{ratioinitial}
\end{eqnarray}
in terms of the energies $e(c_i)$ and $c(e_j)$
of the microscopic configurations $(c_i,c_j)$.

From the full RG rules of Eqs \ref{wijnew} and \ref{picstar},
we obtain the following RG rule for the ratio
\begin{eqnarray}
R^{new}\left({ C}_i  \to { C}_j \right) && = \frac{W^{new} \left({ C}_i  \to { C}_j \right)}
{W^{new} \left({ C}_j  \to { C}_i \right)} 
= \frac{W  \left({ C}_i  \to { C}_j \right)
+ W({ C}_i \to { C}^* ) \times \frac{W \left({ C}^*  \to { C}_j \right)}{W_{out} \left( { C}^*\right)}}
{W  \left({ C}_j  \to { C}_i \right)
+ W({ C}_j \to { C}^* ) \times \frac{W \left({ C}^*  \to { C}_i \right)}{W_{out} \left( { C}^*\right)}} \\
&& = \frac{ \left[ W_{out} \left( { C}^*\right)
W  \left({ C}_j  \to { C}_i \right) \right]
R  \left({ C}_i  \to { C}_j \right)
+ \left[
W({ C}_j \to { C}^* ) W \left({ C}^*  \to { C}_i \right) \right]
R({ C}_i \to { C}^* ) R \left({ C}^*  \to { C}_j \right)}
{\left[
W_{out} \left( { C}^*\right) W  \left({ C}_j  \to { C}_i \right)\right]
+ \left[ W({ C}_j \to { C}^* ) W \left({ C}^*  \to { C}_i \right) \right]}
\nonumber 
\label{ratiorg}
\end{eqnarray}
Let us first consider the first RG step :
since the initial value of Eq. \ref{ratioinitial}
has actually a factorized form, one obtains that the product
of the second term in the numerator is simply
$ R({ C}_i \to { C}^* ) R \left({ C}^*  \to { C}_j \right)
= e^{- \beta (e(c_j)-e(c_i))} = R  \left({ C}_i  \to { C}_j \right)  $.
As a consequence in Eq. \ref{ratiorg}, the factor 
$R  \left({ C}_i  \to { C}_j \right)$ can be factorized 
and what remains in the numerator is exactly equal to the
denominator. This simplification is actually valid at all
steps of the RG procedure by recurrence.
We thus obtained that the ratios $R\left({ C}_i  \to { C}_j \right)$ are actually
'conserved' by the RG flow in the following sense 
\begin{eqnarray}
R^{new}\left({ C}_i  \to { C}_j \right)  = R^{\Gamma=0}\left({ C}_i  \to { C}_j \right)  = e^{- \beta (e(c_j)-e(c_i))}
\label{ratiorgconserved}
\end{eqnarray}
where $c_i$ and $c_j$ are the microscopic configurations that label
the renormalized valleys $C_i$ and $C_j$.

Since the free-energy difference between 
two valleys can be obtained from the ratio of the two
transition rates between them (Eq. \ref{freediffWR}),
we finally obtain 
\begin{eqnarray}
 e^{- \beta  (F(C_j)-F(C_i)) } = R^{new}\left({ C}_i  \to { C}_j \right) =
e^{- \beta (e(c_j)-e(c_i))}
\label{ratiofree}
\end{eqnarray}
i.e. the free-energy difference between the two renormalized valleys
$C_i$ and $C_j$ 
is simply given by the difference of energy of the
two microscopic configurations $c_i$ and $c_j$ that label
the renormalized valleys. 
This property indicates that the statistics of free-energy
differences between renormalized valleys is the same as
the statistics of energy differences between 'good' microscopic
configurations. This property is usually assumed within
the droplet scaling picture \cite{heidelberg,Fis_Hus}
where the low-temperature phase is expected to be governed
by a zero-temperature fixed point.
Within the strong disorder RG procedure, this property emerges 
as a consequence of a conservation law for the ratio
 $R\left({ C}_i  \to { C}_j \right)$
of the transition rates between two renormalized configurations.

As a final remark, we should stress here that this conservation law
does not mean that the free-energy differences remain the same as 
the temperature $T$ varies, because the valleys $C_i$ and $C_j$
that survive during the RG
procedure are temperature-dependent, i.e. the microscopic configurations
 $c_i$ and $c_j$ that label
the longest-lived renormalized valleys will change with the temperature.
Within our present RG framework, the 'chaos properties' with respect
to temperature changes, that are expected within the droplet scaling theory
\cite{heidelberg,Fis_Hus},
should correspond to the sensitivity of the RG flow with respect
to changes in the initial condition, as in Migdal-Kadanoff real-space RG procedures
(see \cite{mckay,Ban_Bray,muriel,thill} and references therein).
However a detailed study of these 'chaos' properties for
the present strong disorder procedure goes beyond the present work
and these effects will not be discussed further here.

\subsection{ Interpretation of the RG rules as an 'adiabatic approximation'  }

\label{adiabatic}

As explained in \cite{Pigo} where the rule of Eq. \ref{wijnew} 
has been proposed to eliminate 'fast states' in dynamical problems 
with two very separated time scales, the prescription to
obtain the new transition rates actually
 corresponds to an 'adiabatic approximation'.
Let us explain this point within our present notations since it has
important consequences for the present the coarse-graining procedure.
 
The idea of the 'adiabatic approximation' \cite{Pigo} 
is that 'fast' degrees of freedom
adapt rapidly to ensure that the probability flow entering into them
is exactly compensated by the probability flow emerging from them.
Within our present notation where the 'fast' state is
the configuration $C^*$, the probability flow entering into $C^*$ 
from its neighbors $(C_1,...,C_n)$ reads
\begin{eqnarray}
J^{in}_{C^*} (t) = \sum_{i=1}^n P_t(C_i) W ( C_i \to C^*)
\label{jin}
\end{eqnarray}
whereas the probability flow emerging from $C^*$
reads
\begin{eqnarray}
J^{out}_{C^*} (t) =  P_t(C^*) W_{out}(C^*) = P_t(C^*) \sum_{i=1}^n  W ( C^* \to C_i)
\label{jout}
\end{eqnarray}
The adiabatic approximation consists in assuming that on time scales
much larger than the typical relaxation time $1/ W_{out}(C^*)$ of $C^*$,
the occupation probability $P_t(C^*)$ adapts to ensure 
the global zero-flow condition $J^{in}_{C^*} (t) = J^{out}_{C^*} (t) $
leading to
\begin{eqnarray}
 P_t(C^*) = \frac{1 }{W_{out}(C^*)} \sum_{i=1}^n P_t(C_i) W ( C_i \to C^*)
\label{probaadiabatique}
\end{eqnarray}
i.e. the occupation probability $P_t(C^*)$ of the 'fast' configuration $C^*$
now only varies slowly in time by following
 the external slow inputs $ P_t(C)$ given
by the 'slow' configurations $C$ that are connected to it.
Physically, this means that the 'fast' configuration $C^*$ is 
equilibrated as much as it can in the 'out-of-equilibrium'
environment produced by the 'slow' configurations.

The equivalence of this way of thinking with the renormalization rule of
Eq \ref{wijnew} is immediate by replacing the value of Eq. \ref{probaadiabatique}
into the evolution equation of a neighbor $C_i$ of $C^*$
\begin{eqnarray}
\frac{dP(C_i)}{dt}
&& =   P_t(C^*) W ( C^* \to C_i) + \sum_{C \neq C^*}
 P_t(C) W ( C \to C_i)- P_t(C_i) W_{out}(C_i)  \\
&& = \left[\frac{1 }{W_{out}(C^*)} \sum_{j=1}^n P_t(C_j) W ( C_j \to C^*) \right]
  W ( C^* \to C_i)+ \sum_{C \neq C^*}
 P_t(C) W ( C \to C_i)- P_t(C_i) W_{out}(C_i) \nonumber \\
&& =  \sum_{C \neq C^*}
 P_t(C) \left[ W ( C \to C_i) 
+ \frac{ W ( C \to C^*) W ( C^* \to C_i) }{W_{out}(C^*)}  \right]
 - P_t(C_i) \left[ W_{out}(C_i) - \frac{ W ( C_i \to C^*) W ( C^* \to C_i) }{W_{out}(C^*)}  \right] \nonumber
\label{equivalence}
\end{eqnarray}
The terms between brackets $\left[... \right]$ 
corresponds exactly to the renormalization
rules of Eqs \ref{wijnew} and \ref{wioutnew}.

\subsection{ RG rule for the energies of valleys  }

\label{rgenergie}

As emphasized in \cite{Pigo}, the 'adiabatic' interpretation shows
that even if one eliminates the 'fast' modes to obtain the effective dynamics
of 'slow' modes, the information on the 'fast' modes is actually not completely
lost since its slow variation in time can be reconstructed via Eq. 
\ref{probaadiabatique} from the dynamics of the 'slow' modes.
Moreover, this point of view allows to derive
how the decimated valley $C^*$ contributes to modify
the energies of the neighboring renormalized valleys.
Equation \ref{probaadiabatique} can be interpreted as follows :
after the elimination of $C^*$, when the system is in the renormalized
configuration $C_i$, it is actually in the decimated configuration $C^*$
with a temporal weight $\frac{W ( C \to C^*) }{W_{out}(C^*)}$.
As a consequence, the appropriate RG rule for the energies of the neighboring
valleys $C_i$
 upon the elimination
of the configuration $C^*$ reads
\begin{eqnarray}
E^{new}(C_i)  = \frac{ E( C_i ) +  E(C^*) \frac{W ( C_i \to C^*) }{W_{out}(C^*)}  }
{ 1 +  \frac{W ( C_i \to C^*) }{W_{out}(C^*)}}
\label{rgener}
\end{eqnarray}
More generally, one may write similar RG equations for observables
that are linear in the probabilities $P_t(C)$, but not for the entropy
which involves logarithms.
(The definition of intra-valleys entropies will be possible only
with the simplified RG rules discussed in section \ref{simplified}).

\subsection{ Discussion on the effects of 'bad decimations' }

The interpretation in terms of an adiabatic
approximation allows also to understand the effects of 'bad' decimations,
 defined as the cases where the exit rate out of
$C^*$ is not very well separated from the exit rates out of its neighbors.
As explained in \cite{Pigo}, there exists some commutativity of the RG rules
in the following sense :  the order of elimination of the 'fast' states
 is not important anymore when all the 'fast' states have been decimated.
The reason is that in the adiabatic approximation,
one may eliminates all the 'fast' states simultaneously 
by requiring that all the corresponding probability currents
vanish simultaneously. The solution of this linear system is unique
and the RG procedure simply corresponds
to solve this system by substitution in a given order,
but any other order would have given the same unique solution.
This shows that when there are only two very different time scales
 as in \cite{Pigo}, all 'fast' states can actually be eliminated
at once.  But in our present framework
for disordered systems where a broad distribution of time scales
exist, from the microscopic time scale of a single move
to the equilibrium time of the whole system, it is not obvious
at the beginning which set of microscopic configurations should be eliminated
to obtain a consistent set of renormalized valleys whose exit rates
is above some prescribed threshold. As a consequence,  
 the iterative RG procedure that renormalizes the exit rates
out of the surviving configurations is actually essential
 to determine which are the longest-lived renormalized valleys  
in each sample. Nevertheless, the 'commutativity' property is important,
since it shows that 'bad decimations'
will actually be corrected in the later stages of the renormalization 
when the neighbors will themselves be decimated.
This phenomenon is very reminiscent of what has been found by D.S. Fisher
for quantum spin chains in Appendix E of \cite{dsf} dedicated to the 
effects of a bad decimation :
' we recover exactly at a later stage from the errors made earlier'.
This property explains why strong 
disorder RG methods are usually much more accurate than one might
think from the approximations involved in the first stages 
of the RG procedure where the probability distributions are not yet
broad enough. The final picture is that at large RG scale $\Gamma$,
 where the probability distributions
are broad enough to ensure a separation of scales between
the decimated configuration and its neighbors :
(i)  the 'present errors' are
small and of order $1/\Gamma$
(ii) the 'past errors' due to the first stages of the RG procedure
have been cured by later decimations.

\subsection{ Example : directed polymer in a two-dimensional medium}

As example of application, we consider the directed polymer
 in a two-dimensional random medium 
(see \cite{Hal_Zha} for a review), first
 introduced to model an interface
 in the low-temperature phase of two-dimensional
disordered ferromagnets \cite{Hus_Hen}. 
 The equilibrium is well described by the Fisher-Huse droplet theory
 \cite{Fis_Hus_DP}
as checked by detailed numerical studies \cite{Fis_Hus_DP,DPexcita}
based on transfer matrix calculations of the partition function
that allow to study big sizes with a good statistics.
Here we wish to follow numerically the strong disorder RG procedure
to obtain the properties of the longest-lived valleys.
Since we work in configuration space, we will not be able to study big
system sizes (see \cite{rglong} for more details),
 but our aim is to see that the strong disorder RG procedure can be applied
consistently and yields appropriate results for the free-energy difference
and for the weight statistics of the two longest-lived valleys. 
The numerical details are described in our previous works
on the non-equilibrium dynamics \cite{rgshort,rglong},
where we have checked the convergence towards an infinite disorder fixed
point for the distribution of the renormalized barriers and where
we have studied the distribution over samples of the equilibrium time.
Here we present our results for the observables characterizing the valleys 
weights.

\begin{figure}[htbp]
\includegraphics[height=6cm]{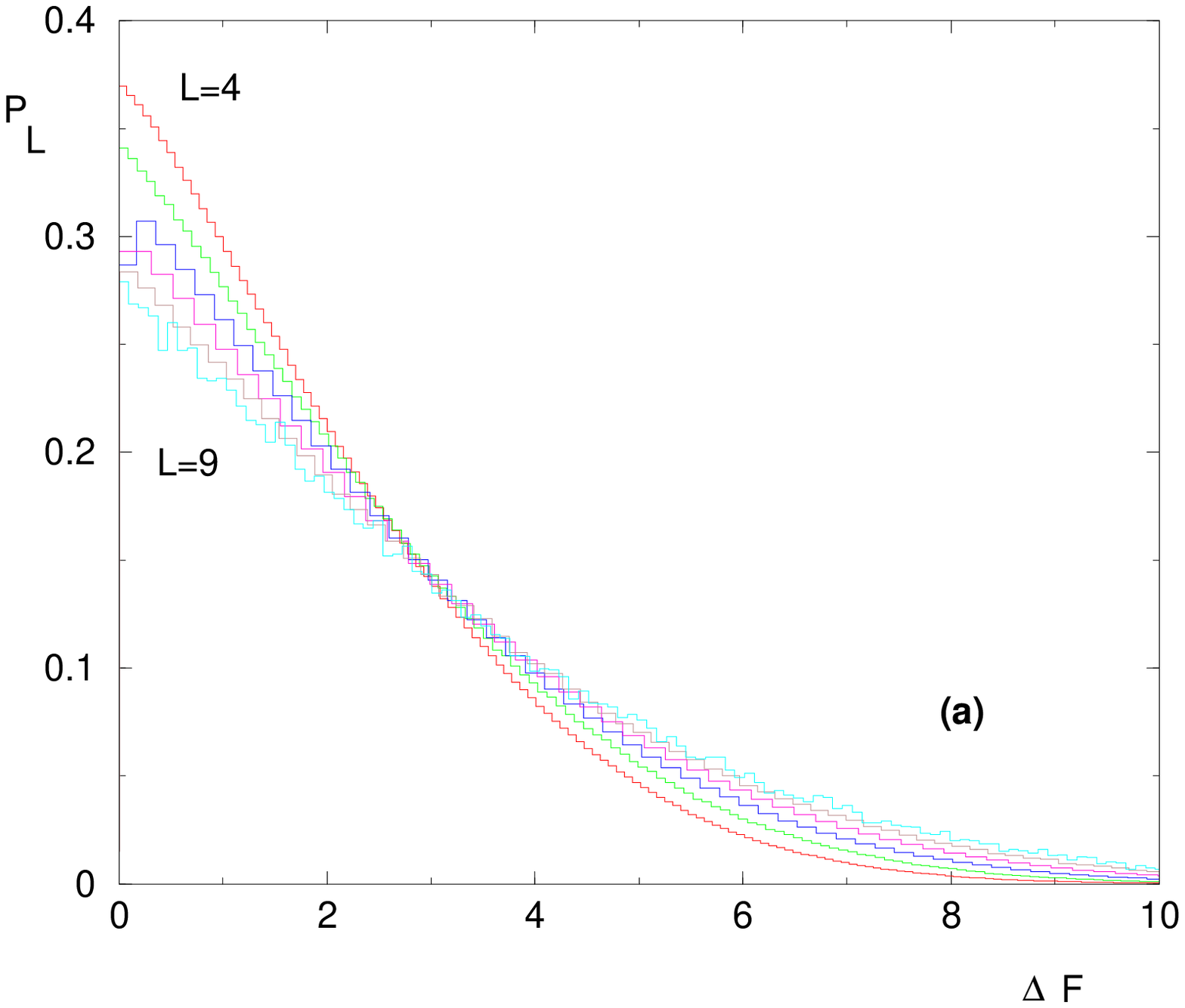}
\hspace{1cm}
\includegraphics[height=6cm]{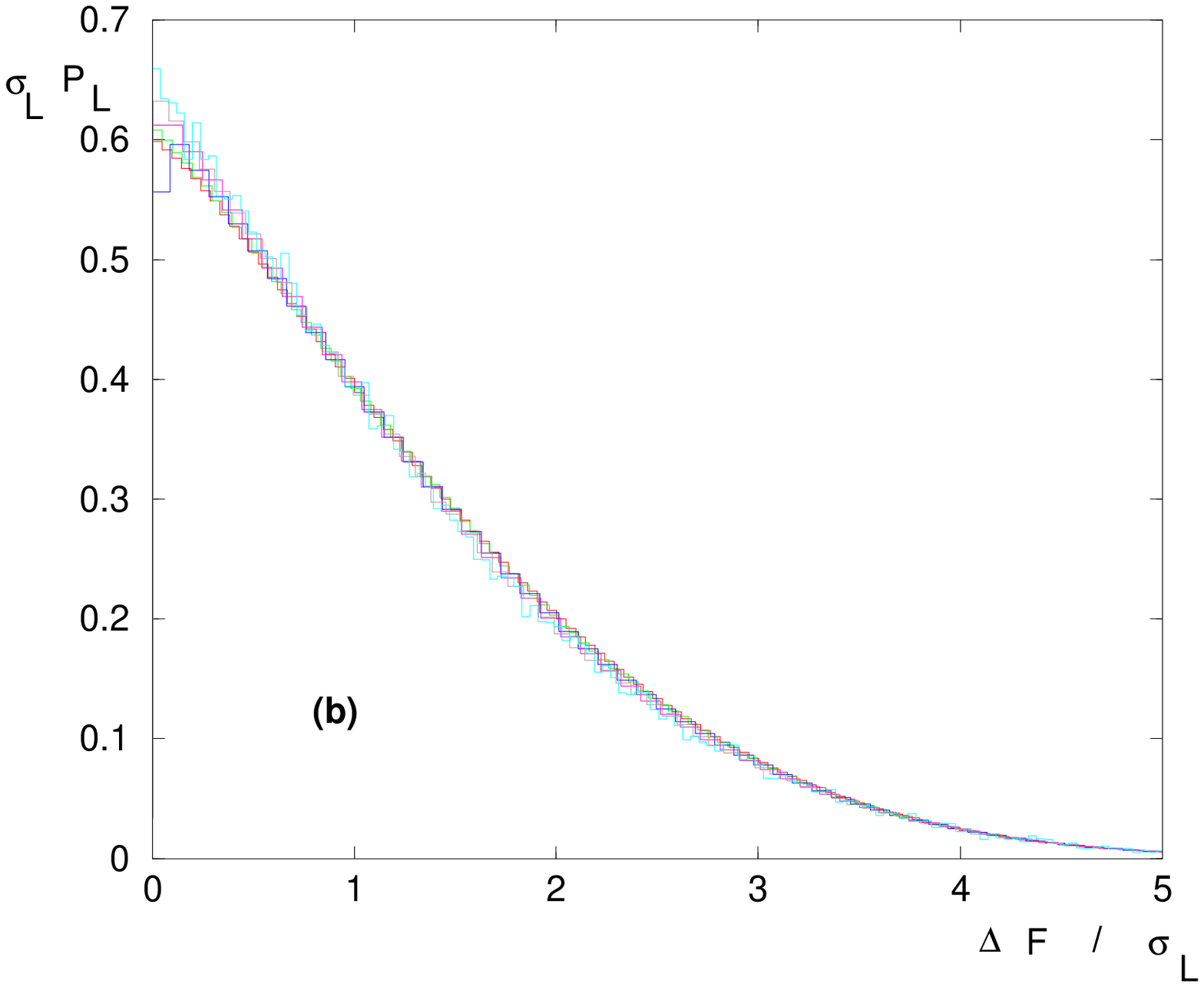}
\caption{ Statistics over the samples of the free-energy difference
 $\Delta F =F_2-F_1$ between the two last valleys of a disordered sample
(a) Probability distribution $P_L(\Delta F)$ for the sizes $L=4,5,6,7,8,9$
 ( corresponding to $2^L \leq 512$ initial configurations )
with a statistics of $n_s \geq 10^5$ disordered samples
(we have data up to $L=11$ with $n_s=1200$ samples, 
but histograms are too noisy). 
(b) same data after the rescaling by the width $\sigma_L$ :
 $\sigma_L P_L$ as a function of $\Delta F / \sigma_L$
(see Eq. \ref{rescalfree}). }
\label{fighistofreergthermo}
\end{figure}

\subsubsection{ Histogram of the free-energy difference between
the two last valleys }

For each disordered sample of size $L$, we obtain via the strong
disorder renormalization the free-energy difference
 $\Delta F =F_2-F_1$ between the two last valleys
(see Eq \ref{freediffWR} and \ref{ratiofree}).
We show on Fig. \ref{fighistofreergthermo} (a)
the probability distribution $P_L(\Delta F)$ over the samples
of this free-energy difference for various sizes.
As expected from the droplet scaling picture \cite{heidelberg,Fis_Hus},
this probability distribution can be rescaled by its width $\sigma_L$
\begin{eqnarray}
P_{L}( \Delta F) \sim  
  \frac{1}{\sigma_L } {\tilde P} 
\left( u \equiv \frac{ \Delta F }{ \sigma_L }
 \right)
\label{rescalfree}
\end{eqnarray}
and the rescaled distribution ${\tilde P}$ is finite at the origin
${\tilde P}(0)>0$ : the rescaling of our data shown on
 Fig. \ref{fighistofreergthermo} (b) corresponds exactly to the
expected shape for
droplet free energies distribution (see Fig. 2 of Ref. \cite{Fis_Hus}).
Of course the sizes studied here $4 \leq L \leq 11$
are too small to obtain from the width $\sigma_L \sim L^{\theta}$
 a precise measure of 
 the droplet exponent which is exactly known to be
$\theta=1/3$ in the present case \cite{Hus_Hen}.

\subsubsection{ Histograms of $S^{inter}$ and $Y_2$ for the two last valleys }

\begin{figure}[htbp]
\includegraphics[height=6cm]{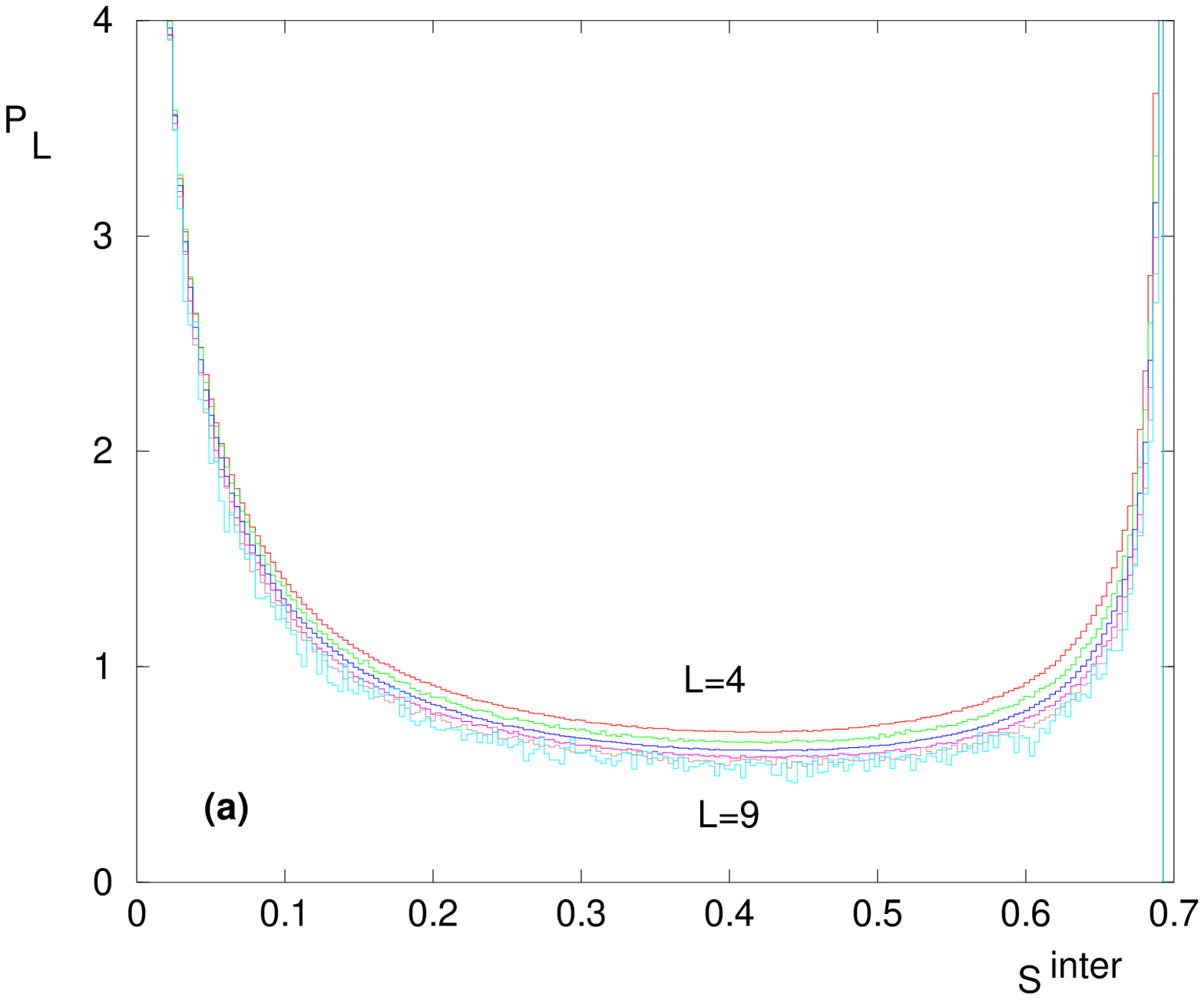}
\hspace{1cm}
\includegraphics[height=6cm]{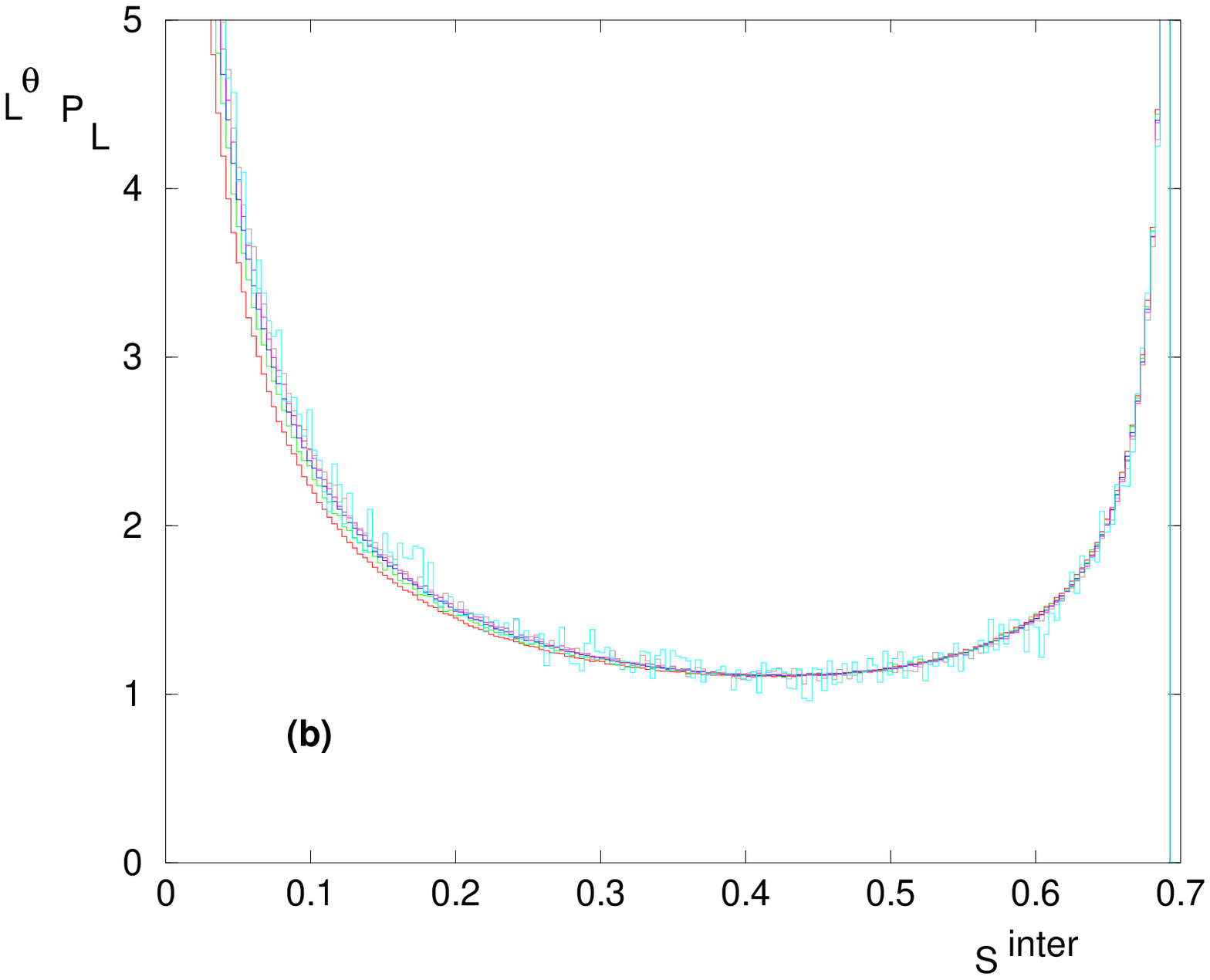}
\caption{ Statistics of the intervalley
entropy $S^{inter}= -p \ln p -(1-p) \ln (1-p) $
for the two
last valleys with $p= \frac{1}{1+e^{-\beta \Delta F}}$
 (a) Probability distribution $P_L(S^{inter})$ for the sizes $L=4,5,6,7,8,9$
 (b) Same data after the rescaling : $L^{\theta} P_L(S^{inter})$ 
with $\theta=1/3$. }
\label{figsinterrgthermo}
\end{figure}

\begin{figure}[htbp]
\includegraphics[height=6cm]{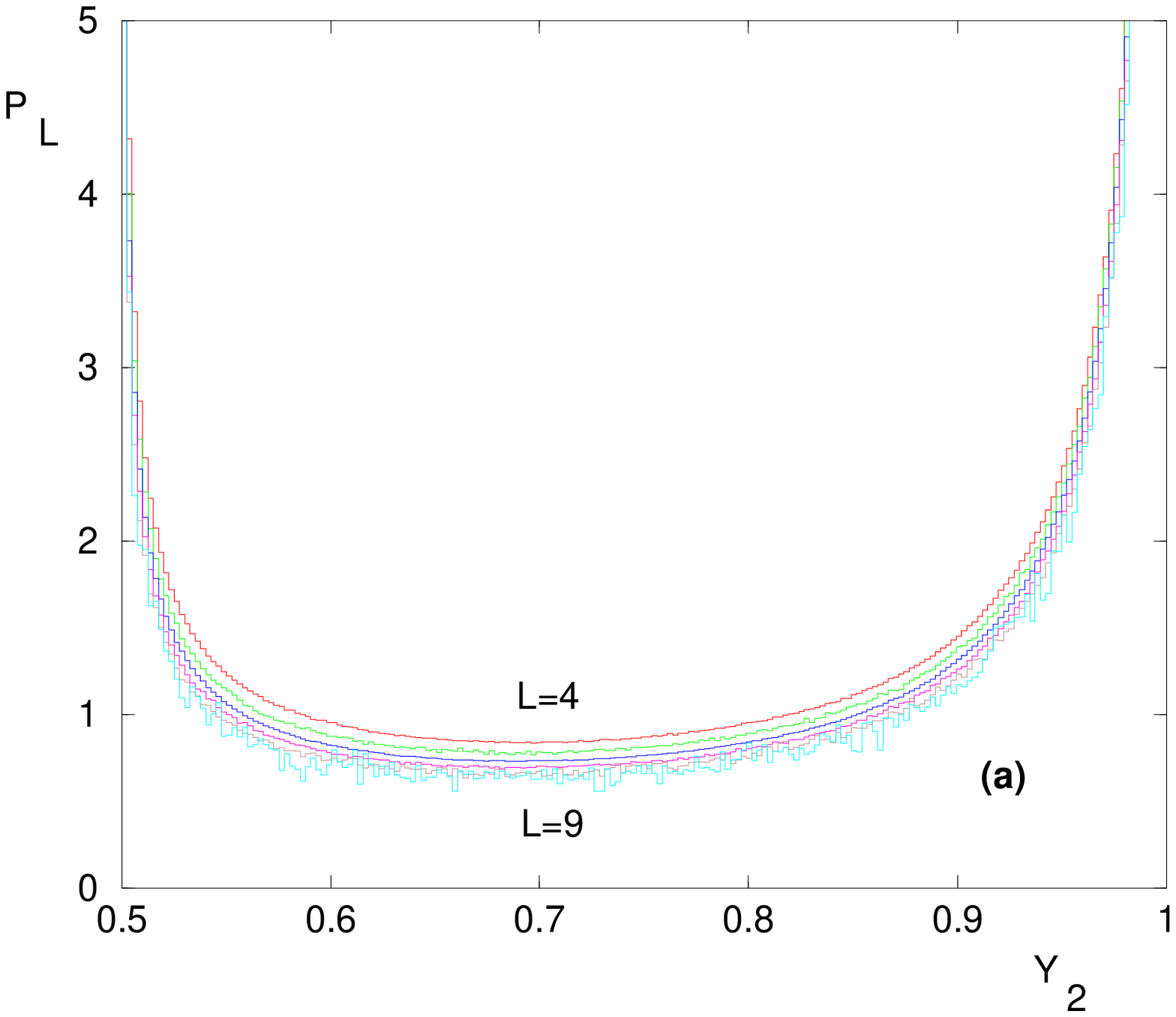}
\hspace{1cm}
\includegraphics[height=6cm]{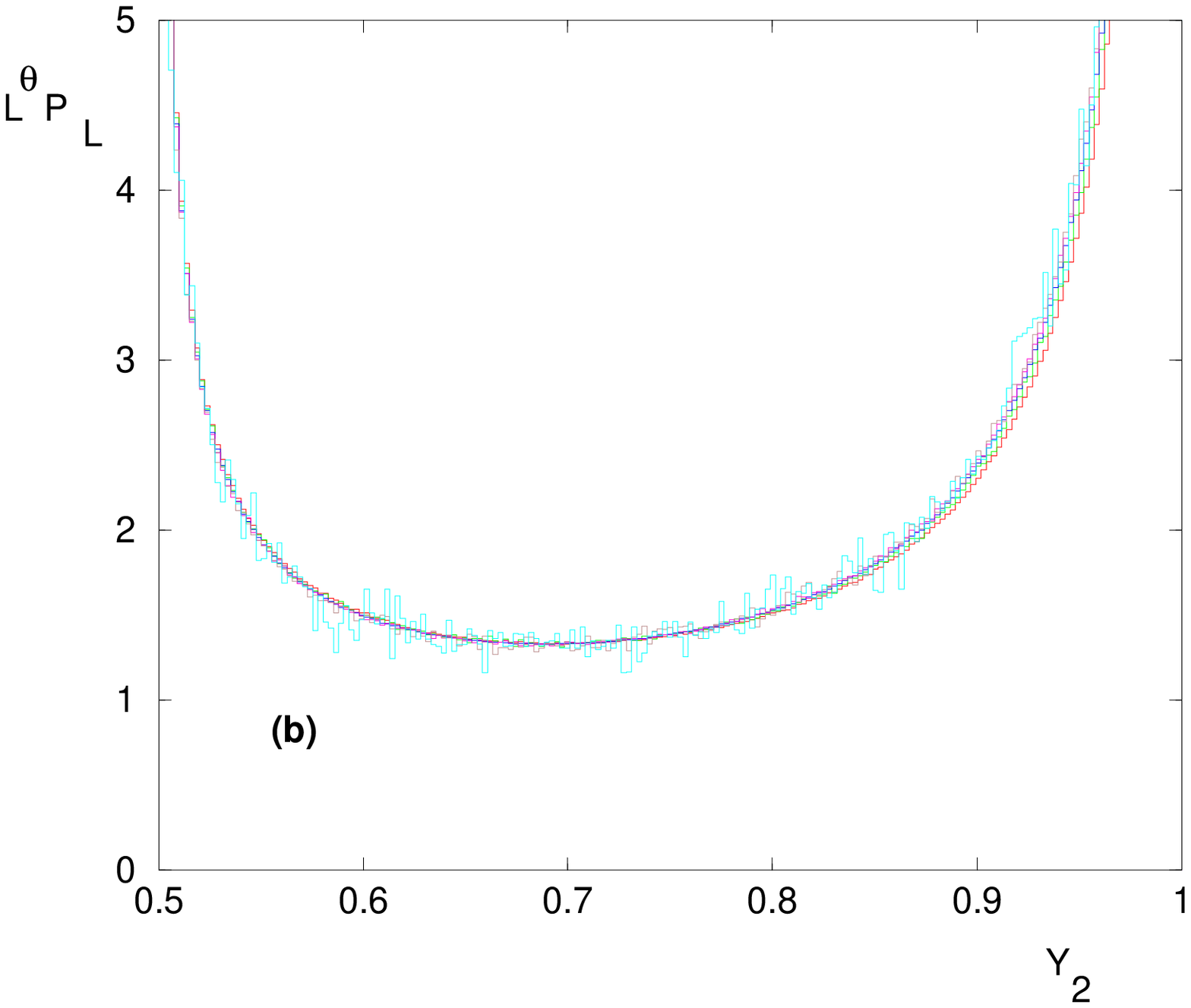}
\caption{ Statistics of $Y_2=p^2+(1-p)^2$ for the two
last valleys with $p= \frac{1}{1+e^{-\beta \Delta F}}$
 (a) Probability distribution $P_L(Y_2)$ for the sizes $L=4,5,6,7,8,9$ 
 (b) Same data after the rescaling : $L^{\theta} P_L(Y_2)$
with $\theta=1/3$. }
\label{figy2rgthermo}
\end{figure}

The free-energy difference $\Delta F$ between the two last valleys
determines their respective weights $(p,1-p)$ in configuration space
with
\begin{eqnarray}
p = \frac{1}{1+e^{-\beta \Delta F}}
\label{pweight}
\end{eqnarray}
We show on Fig. \ref{figsinterrgthermo} and \ref{figy2rgthermo} 
the histograms of the corresponding inter-valley entropy (Eq. \ref{sinter})
\begin{eqnarray}
 S^{inter}= - p \ln p -(1-p) \ln (1-p) 
\label{sinterbis}
\end{eqnarray}
and of the second generalized moment of Eq. \ref{defyk}
\begin{eqnarray}
 Y_2=  p^2+(1-p)^2
\label{defy2}
\end{eqnarray}
for various sizes.
For a positive droplet exponent $\theta>0$, 
one expects that the corresponding distributions
converge respectively towards $\delta(S^{inter})$ and to
$\delta(Y_2-1)$, with corrections of order $1/L^{\theta}$
on the intervals $0<S^{inter}<\ln 2$ and $1/2<Y_2<1$.
 We thus show on Fig. \ref{figsinterrgthermo} b
 and \ref{figy2rgthermo} b the rescaled data by a factor $L^{1/3}$ :
we observe that the data collapse is better in the region of nearly degenerate
weights between the two valleys,
 corresponding to $S^{inter}$ near $\ln 2$ and to $Y_2$ near 1/2.
This seems to indicate that the corrections to the asymptotic scaling
with the droplet exponent $\theta=1/3$ are smaller in the region 
of small free-energy difference
 than in the region of large free-energy difference.


\section{  'Simplified' RG rules near infinite disorder fixed points  } 

\label{simplified}

This section recalls the 'simplified RG rules' introduced
  in \cite{rglong} and discuss their consequences for the valleys properties,
in terms of quasi-equilibrium within each renormalized valleys.

\subsection{ Reminder on the simplified RG rules 
based the preferred exit channel \cite{rglong} } 

Whenever the flow is towards some `infinite-disorder' fixed point,
where the distribution of renormalized exit barriers becomes
 broader and broader
upon iteration (Eqs \ref{pgammabout} and \ref{infinite}),
one expects that the exit rate
out of the decimated configuration ${ C}^*$  
\begin{eqnarray}
W_{out} \left( { C}^* \right)  =
 \sum_{i=1}^n  W \left({ C}^* \to  { C}_i \right) 
\label{woutcstar}
\end{eqnarray}
will actually be dominated by the preferred exit channel
$i_{pref}$ having the largest contribution in the sum of Eq. \ref{woutcstar} 
\begin{eqnarray}
W_{out} \left( { C}^* \right)  \simeq
  W \left({ C}^* \to  { C}_{i_{pref}} \right) 
\label{wpref}
\end{eqnarray}
i.e. one expects that the probability distribution 
$\pi_{{ C}^*} \left({ C}_j \right)$ of Eq. \ref{picstar}
will become a delta distribution on the preferred exit channel
up to exponentially small terms 
\begin{eqnarray}
\pi_{{ C}^*} \left({ C}_j \right) \simeq \delta_{j,i_{pref}}
 +...
\label{picstardelta}
\end{eqnarray}

 The dominance of the preferred exit channel (Eq. \ref{wpref})
near an infinite disorder fixed point has been checked numerically
in \cite{rglong} for the case of the directed polymer in a two-dimensional
random medium. However, we expect that
it holds more generally for the following reasons.
The RG rules with their characteristic multiplicative structure
of Eqs \ref{wijnew} and \ref{picstar}
act directly on the transition rates 
$ W \left({ C}_j \to  { C}_i \right) $ between configurations,
whereas the total exit rates $W_{out}$ are derived quantities
obtained by summing over the possible exit channels. The notion
of convergence towards an infinite disorder fixed point
has been defined above by the property that 
the probability distribution of the remaining exit rates $W_{out}$
becomes broader and broader. However we expect that when it happens,
it is because the probability distribution of the individual
transition rates 
$ W \left({ C}_j \to  { C}_i \right) $ themselves
 becomes broader and broader (see \cite{rglong} for more details).

The 'simplified RG procedure' introduced in \cite{rglong}
can be summarized as follows :

(1) the first point is the same as in the 'full RG rules' ( Eq \ref{defwmax})

(2') among the neighbors $({ C}_1,{ C}_2,...,{ C}_n)$
 of configuration ${ C}^*$, find the preferred exit channel $i_{pref}$.
Update the
transitions rates from the $(n-1)$ non-preferred
neighbors $i \ne i_{pref}$ towards $i_{pref}$ by
the approximated rule 
\begin{eqnarray}
W^{new}({ C}_i \to { C}_{i_{pref}} ) \simeq
W({ C}_i \to { C}_{i_{pref}} )
+ W({ C}_i \to { C}^* )
\label{wijnewtowardsipref}
\end{eqnarray}
where the probability distribution $\pi_{{ C}^*} \left({ C}_j \right)$ of the full RG rule of Eqs \ref{wijnew} and \ref{picstar}
has been replaced by the leading delta function of Eq. \ref{picstardelta}.
Update the transitions rates from $i_{pref}$ towards
the $(n-1)$ non-preferred neighbors $i \ne i_{pref}$ by the full RG rule
of Eqs \ref{wijnew} and \ref{picstar}
\begin{eqnarray}
W^{new}({ C}_{i_{pref}}\to { C}_i ) =
W({ C}_{i_{pref}} \to { C}_i )
+ W({ C}_{i_{pref}} \to { C}^* ) \times 
\frac{W \left({ C}^*  \to { C}_i \right)}
{W_{out} \left( { C}^*\right)}
\label{wijnewfromipref}
\end{eqnarray}
Here the full rule is used because the ratio 
$\frac{W \left({ C}^*  \to { C}_i \right)}
{W_{out} \left( { C}^*\right)}$ is small and should thus
be evaluated correctly.

(3') With the rule of Eq. \ref{wijnewtowardsipref}, the exit rates out of
the $(n-1)$ non-preferred neighbors $i \ne i_{pref}$ do not have to be updated
since the exit rate towards 
 ${ C}^*$ has been completely transfered to $i_{pref}$.
So the only update of exit rate is for the preferred neighbor $i_{pref}$
via the definition of Eq. \ref{wioutnewactualisation}
or with the equivalent rule of Eq. \ref{wioutnew}.

(4) return to (1)

As explained in \cite{rglong}, these simplified RG rules 
are interesting both 
from a computational point of view (they allow to study bigger system sizes)
and from a theoretical point of view 
to make the link with the idea of 'internal ergodicity'
in each valley \cite{Palmer} as we now explain.

\subsection{ Interpretation in terms of quasi-equilibrium 
within valleys} 

\label{metastable}

In the studies on slowly relaxing systems such as disordered systems,
glasses or granular media, it is usual to separate
the dynamics into two parts : there are `fast' degrees of freedom
which rapidly reach local quasi-equilibrium plus a slow non-equilibrium part.
Within the present strong disorder renormalization in configuration space,
these ideas can be applied directly as follows.
To each time $t$, one may associate a set of valleys
which are labelled by the surviving configurations at the
RG scale $\Gamma= \ln t$. 
Since at large scale, the RG flow for the barrier distribution
is towards some ``infinite disorder'' fixed 
point, the different time scales are effectively very well separated.
And thus asymptotically we recover the Palmer's decomposition 
into 'components' from which the escape probability is small
and in which there is 'internal ergodicity', i.e.
the interior of each valley has been able to equilibrate
\cite{Palmer,Palmer83}.
Moreover, the slow non-equilibrium part of the dynamics corresponds to the 
evolution of the renormalized valleys with the RG scale :
some valleys disappear and are absorbed by a neighboring valley.
Here again, we see that the RG procedure corresponds asymptotically
to Palmer 'bifurcation cascade of components' :
see Figure 4 of \cite{Palmer} where the 'bifurcation cascade'
is drawn as a function of temperature for a fixed time.
In our present framework, it is better to represent 
this 'bifurcation cascade' as a function of time for fixed temperature.
This is because it has been understood since the papers of Palmer
 \cite{Palmer,Palmer83} that disordered systems have some 'chaos'
property with respect to temperature changes, i.e. 
the valleys that will emerge at large scales for different temperatures
are not simply related (for more details on these chaos properties,
see \cite{heidelberg,Fis_Hus,Ban_Bray,muriel,thill}).

To finish this discussion, we would like to emphasize a very
important point : the asymptotic dominance of the preferred exit channel
 near the infinite disorder
fixed point is actually crucial to obtain quasi-equilibrium
inside valleys.
In particular, if the degeneracy between the second preferred exit channel 
and the first preferred exit channel could occur with a finite probability,
then finite contributions of out-of-equilibrium situations
at all scales would ruin the quasi-equilibrium approximation 
 : the probability to be in a configuration ${ C}$
at time $t$ would not depend only on its energy $E({ C})$ 
and on the partition function of the renormalized valley
it belongs to, but would be instead a very complicated function
of all possible paths from ${ C}_0$ to ${  C}$ 
with their appropriate dynamical weights.
To better understand the importance of this discussion,
it is useful to recall here a well-identified
 exception of the quasi-equilibrium idea, namely
  the symmetric Bouchaud's trap model in one dimension,
where even in the limit of arbitrary low temperature,
the diffusion front in each sample consists
 in two delta peaks, which are completely out of equilibrium with each other
\cite{traprg} : the weights of these two delta peaks
do not depend on their energies, but instead 
 on the distances to the origin
that determine the probability to reach one before
the other (see \cite{traprg} for more details).
In this trap model, the reason is clear : whenever the particle
escapes from a trap, it jumps either to the right or to the left
with equal probabilities $(1/2)$, i.e. the two possible exit channels
are degenerate by the very definition of the model that imposes
this symmetry. In other disordered models where this degeneracy is not
imposed by a symmetry of the model, this degeneracy can only 
occur accidentally with some probability. The question is then whether
this probability of accidental degeneracy between
the two preferred exit channels remains finite or 
becomes rare (i.e. decays to zero) at large times. 
Within the present strong disorder RG where the flow is towards some
infinite disorder fixed point, the dominance of the preferred exit channel
precisely means that the probability of these accidental degeneracy
decays to zero, so that the quasi-equilibrium approximation 
 becomes asymptotically exact at large times.
(As discussed in detail in \cite{sinaimetastablestates}
for the case of the Sinai model and in \cite{rglong} for
our present RG procedure in configuration space,
the rare events where the quasi-equilibrium approximation breaks down
 occur with a vanishing probability of order
$1/\Gamma=1/(\ln t)$ at large times.)

\subsection{ RG rules for the intra-valley energies and entropies   } 

The property of quasi-equilibrium inside each valley discussed above
allows to write RG rules for the intra-valley energies and entropies 
as we now explain.
Within the simplified RG rules where the preferred exit channel $C_{i_{pref}}$
out of the decimated configuration $C^*$ actually dominates,
the decimation of $C^*$ can be interpreted as the merging
of the two valleys $C_{i_{pref}}^{old}$ and $C^*$ into
a single quasi-equilibrated valley 
$C_{i_{pref}}^{new}=C_{i_{pref}}^{old} \cup C^*$.
 The ratio of the times spent in the two sub-valleys $C^*$
and $ C_{i_{pref}}^{old}$ once they are at equilibrium with each other
is given by the ratio of the renormalized transition rates
\begin{eqnarray}
\lambda \equiv \frac{p_{temporal}({ C}^*)}{p_{temporal}({ C}_{ipref}^{old})}
\simeq \frac{W \left({ C}^{ipref}_{old}  \to { C}^* \right)}
{W \left({ C}^*  \to { C}^{ipref}_{old} \right)}
\label{deflambda}
\end{eqnarray}
Since we have defined the free-energies differences between valleys
from this ratio of the renormalized transition rates (Eq. \ref{freediffWR}),
one has
\begin{eqnarray}
\lambda = e^{- \beta (F(C^*)-F({ C}_{ipref}^{old}))} 
\label{lambdafree}
\end{eqnarray}
i.e. the relative weights of the two sub-valleys
is related to the free-energy difference of the two sub-valleys.

The full RG rule of Eq. \ref{rgener} for the energies of the renormalized
valleys in contact with $C^*$ reduces, within the simplified RG rules, to
the following renormalization for the energy of the valley $C_{i_{pref}}$ 
\begin{eqnarray}
E(C^{new}_{i_{pref}})  = \frac{ E( C^{old}_{i_{pref}} ) +  \lambda E(C^*)  }
{ 1 +  \lambda}
\label{rgenersimpli}
\end{eqnarray}
Again, the physical meaning of this rule is very clear :
 the two sub-valleys are now
 at equilibrium with each other inside the bigger valley.

Within this quasi-equilibrium picture, it becomes possible
to define an intra-valley entropy $S^{intra}(C)$,
and to write the corresponding RG rule when the valley $C^*$ is decimated.
With the normalization 
$p_{temp}({ C}^*)+p_{temp}({ C}^{ipref}_{old})=1$
and the ratio of Eq. \ref{deflambda}, the weights of
the two sub-valleys are respectively
$p_{temp}({ C}^*)=\lambda/(1+\lambda) $ and 
$p_{temp}({ C}^{ipref}_{old})=1/(1+\lambda) $.
Now taking into account that each sub-valley has its own internal entropy,
one obtains that the internal entropy $S^{intra}$
(defined as $S^{intra}= -\sum p(c) \ln p(c)$ in terms of the weights $p(c)$
of microscopic configurations $c$ that belongs to the valley)
evolves according to
\begin{eqnarray}
S^{intra}(C^{new}_{i_{pref}}) = \frac{  S^{intra} ( C^{old}_{i_{pref}} )
+  \lambda S^{intra}(C^*)   }
{  1+ \lambda  } + \left[
\ln (1+ \lambda) -\frac{\lambda}{1+ \lambda} \ln \lambda \right]
\label{rgs}
\end{eqnarray}
The first term comes from the internal entropies of the sub-valleys,
whereas the second term between $[...]$ corresponds
to the merging-entropy of the two sub-valleys
$S_{merging}= - q \ln q -(1-q) \ln (1-q)$ with $q=1/(1+\lambda)$.

The valley free-energy 
that can be defined only from the inter-valley dynamics within the full RG rules
can be also expressed in terms of the 'interior' of the valley
within the simplified RG rules,
because now upon decimation, the decimated valley $C^*$ is attributed to a single 
surviving valley $C_{ipref}$. 
The RG rule for the free-energy simply correspond to the sum of the partition
functions of the two sub-valleys
\begin{eqnarray}
e^{- \beta F(C^{new}_{i_{pref}})} =
e^{- \beta F(C^{old}_{i_{pref}})} + e^{- \beta F(C^*)} 
= e^{- \beta F(C^{old}_{i_{pref}})} 
\left[ 1+  e^{- \beta (F(C^*)-F({ C}_{i_{pref}}^{old}))}  \right]
\label{rgzintra}
\end{eqnarray}
i.e. using the definition of $\lambda$ (Eq. \ref{lambdafree}), one obtains
\begin{eqnarray}
F(C^{new}_{i_{pref}}) =  F(C^{old}_{i_{pref}}) -T \ln (1+\lambda)
\label{rgfreeintra}
\end{eqnarray}
As it should for consistency, the RG equations for the energy, the entropy
and the free-energy are compatible with the thermodynamic relation
$F=E-T S^{intra}$ inside each valley.

\subsection{ Numerical results for the
 directed polymer in a two-dimensional medium}

The numerical details to follow the 'simplified RG rules'
for the directed polymer in a two-dimensional medium
are described in our previous works
on the non-equilibrium dynamics \cite{rgshort,rglong},
where we have checked (i) the convergence towards an infinite disorder fixed
point for the distribution of the renormalized barriers 
(ii) the validity of the simplified RG rules 
by comparing the rescaled distributions obtained via the full RG rules
and via the simplified RG rules,
and where we have measured the barrier exponent $\psi$ from
the distribution over samples of the equilibrium time.
Here we present our results for the observables characterizing the valleys.

\subsubsection{ Statistics of the free-energy difference between
the two last valleys }

\begin{figure}[htbp]
\includegraphics[height=6cm]{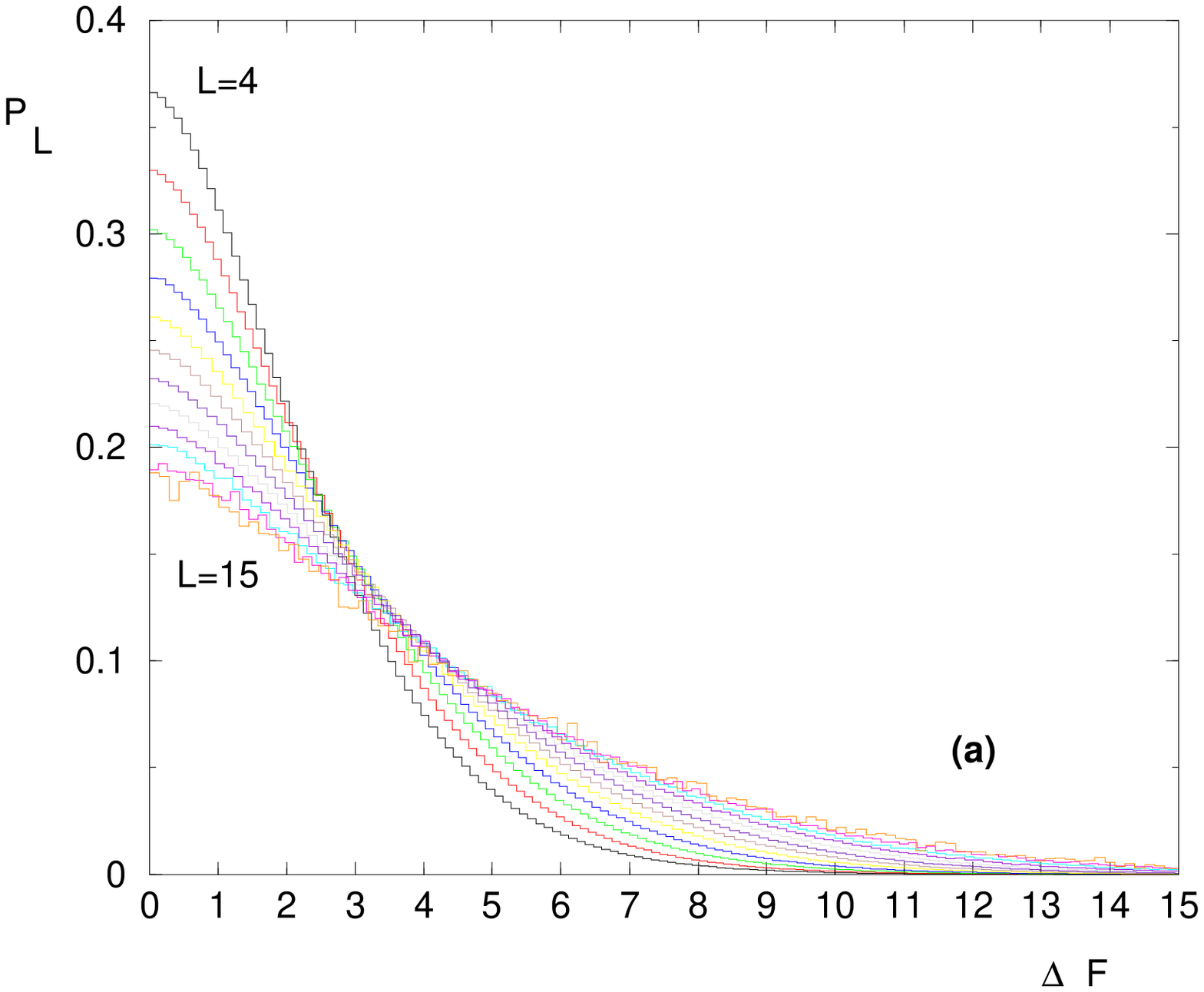}
\hspace{1cm}
\includegraphics[height=6cm]{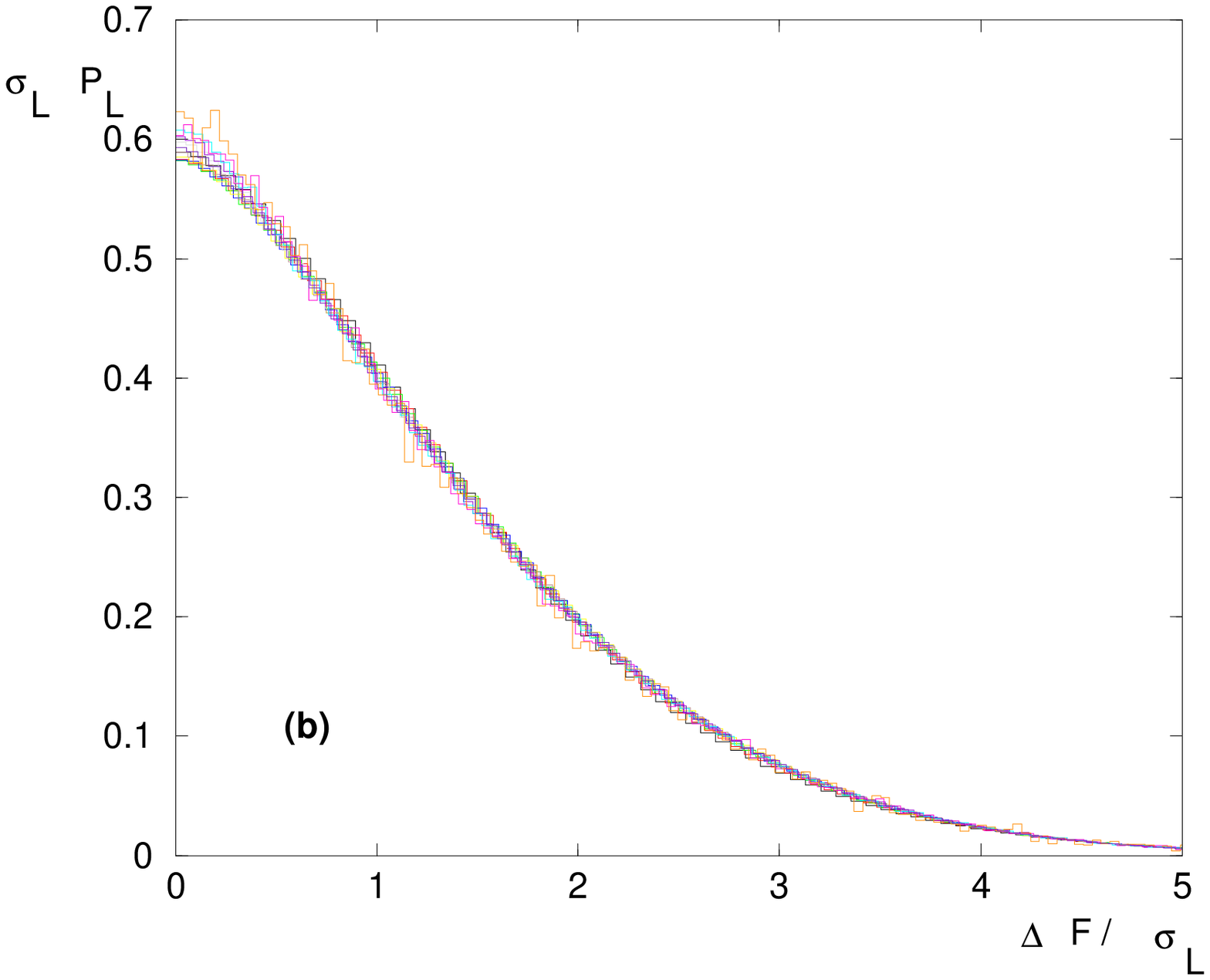}
\caption{ Statistics over the samples of the free-energy difference
 $\Delta F =F_2-F_1$ between the two last valleys of a disordered sample
using the simplified RG rules
(a) Probability distribution $P_L(\Delta F)$ for the sizes $4 \leq L \leq 15$
 ( corresponding to $2^{L} \leq 32768$ initial configurations )
with a statistics of $n_s \geq 5.10^4$ disordered samples
(we have data up to $L=18$ with $n_s=800$ samples, 
but histograms are too noisy). 
(b) same data after the rescaling by the width $\sigma_L$ :
 $\sigma_L P_L$ as a function of $\Delta F / \sigma_L$. }
\label{fighistofreergthermosimpli}
\end{figure}

We show on Fig. \ref{fighistofreergthermosimpli} the probability distribution
$P_L(\Delta F)$ of the free-energy difference
 $\Delta F =F_2-F_1$ between the two last valleys of a disordered sample
obtained via the simplified RG rules.
The rescaled distribution shown on Fig. \ref{fighistofreergthermosimpli} (b)
coincides with the one obtained with the full
RG rules (see Fig. \ref{fighistofreergthermo} b ). This shows 
that the simplified RG rules capture correctly
the fixed point properties of the valleys.

\subsubsection{ Statistics of the entropy
 difference and energy difference between
the two last valleys }

\begin{figure}[htbp]
\includegraphics[height=6cm]{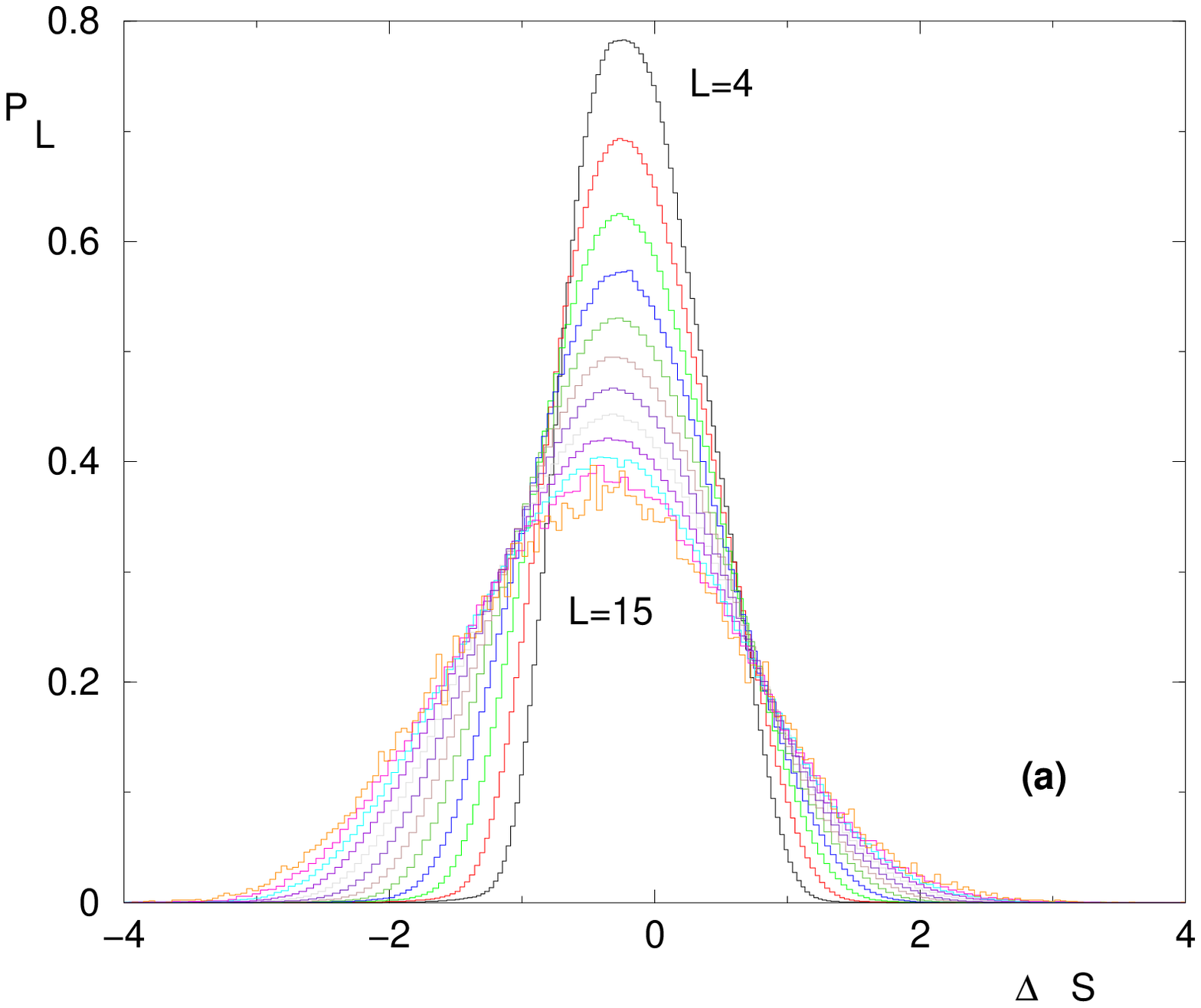}
\hspace{1cm}
\includegraphics[height=6cm]{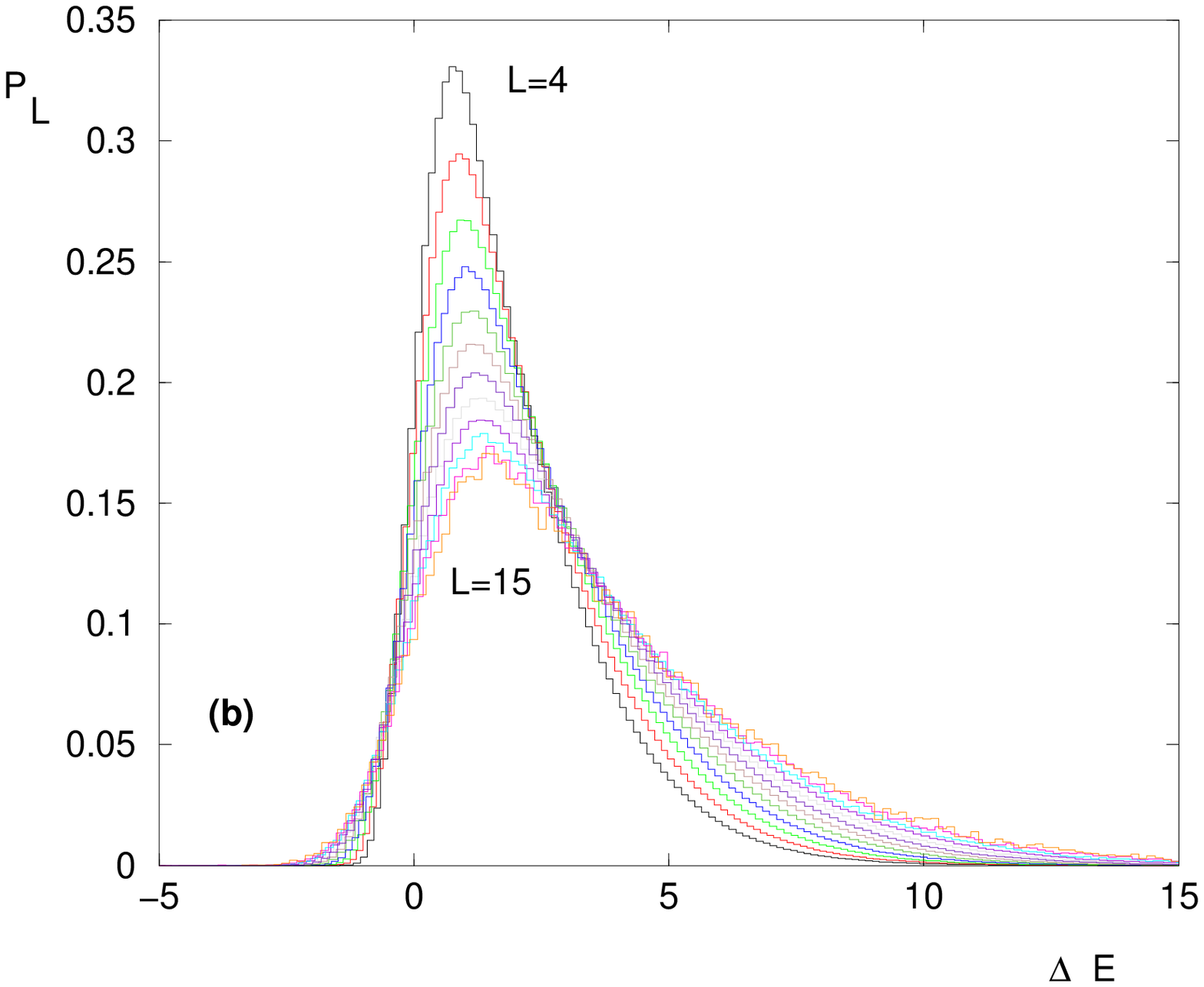}
\caption{ Statistics over the samples of the entropy difference
 $\Delta S =S^{intra}_2-S^{intra}_1$ and of the energy difference
 $\Delta E =E_2-E_1$ between the two last valleys of a disordered sample
(a) Probability distribution $P_L(\Delta S)$ for the sizes $4 \leq L \leq 15$ :
these distribution converge towards a Gaussian shape.
(b) Probability distribution $P_L(\Delta E)$ for the sizes $4 \leq L \leq 15$. }
\label{figentropieenergiergthermosimpli}
\end{figure}

We show on Fig. \ref{figentropieenergiergthermosimpli} (a)
the probability distribution $P_L(\Delta S)$ of the entropy difference
 $\Delta S =S^{intra}_2-S^{intra}_1$ 
between the two last valleys of a disordered sample :
these distribution converge towards a Gaussian shape, in agreement with
the droplet scaling theory where the entropy is dominated by 
a sum of independent small-scale contributions \cite{Fis_Hus}.
One then expects that the entropy difference has the
Central-Limit scaling 
\begin{eqnarray}
\Delta S_L \simeq L^{1/2} u
\label{entropydemi}
\end{eqnarray}
where $u$ is a Gaussian variable.
With our data, we observe the Gaussian distribution, 
but the sizes $4 \leq L \leq 18$ are too small to measure precisely the
exponent $1/2$.

We show on Fig. \ref{figentropieenergiergthermosimpli} (b)
the probability distribution $P_L(\Delta E)$ of the energy difference
 $\Delta E =E_2-E_1$ between the two last valleys of a disordered sample  :
as a consequence of the thermodynamic relation $E=F+TS$ inside each valley,
the distribution of the energy difference 
is a convolution of the free-energy distribution of Fig. 
\ref{fighistofreergthermosimpli} (a) with the gaussian entropy distribution
of Fig. \ref{figentropieenergiergthermosimpli} (a).
Our conclusion is thus that the distributions of
the free-energy difference (Eq. \ref{rescalfree}) and of the entropy
difference (Eq. \ref{entropydemi}) 
are the two primary distributions that have good scaling properties
even on the small sizes considered here, whereas the
probability distribution of the energy is a mixture of these two.

\section{ Summary and conclusions  }

\label{conclusion}

From the point of view of 'broken ergodicity' \cite{Palmer,Palmer83}
that we have adopted in this paper to describe
the equilibrium properties of disordered systems,
the appropriate valleys 
are defined as the 'components' of configuration space
that are separated by large barriers, i.e. the valleys are defined with respect
to their capacity to 'confine' the dynamics \cite{Palmer,Palmer83}. 
In the present paper we have explained how
the strong disorder renormalization procedure allows to construct
in each sample
 all the valleys that are separated by barriers greater than a
prescribed threshold $\Gamma$ representing the RG scale.
To make the link with previous approaches,
we have discussed in details the physical interpretations 
of the 'full RG' rules and of the 'simplified RG' rules
in terms of the 'adiabatic approximation' \cite{Pigo}
and in terms of the 'quasi-equilibrium' 
inside each valley \cite{Palmer} respectively.
We have also explained how this explicit RG formulation gives new insights
into the main ingredients of the droplet scaling picture.
In particular,  the 'zero-temperature' nature of the fixed point for
the probability distribution of the free-energy difference between valleys,
emerges here from a special conservation law along the RG flow
for ratios of renormalized transition rates.
As an example of application, we have followed numerically
the strong disorder RG rules for the directed polymer
in a two dimensional random medium to obtain the statistical properties
of the free-energy difference, the entropy difference
and the energy difference between the two longest-lived valleys.
In particular, we have obtained that the distribution of
the entropy difference converges towards a Gaussian in agreement
with the droplet scaling theory where the entropy is dominated
by independent small scales contributions.
Our conclusion is that an excitation has two independent
 primary properties which are
(i) its free-energy $\Delta F \sim L^{\theta} v$ that involves the droplet
exponent ($\theta=1/3$ for the directed polymer 
in a two dimensional random medium)
(ii) its entropy  $\Delta S \sim L^{d_s/2} u$ that involves
the dimensionality $d_s$ of the surface of an excitation
($d_s=1$ for the directed polymer).
These two quantities present nice scaling
behaviors even on moderate system sizes, whereas the energy 
$\Delta E=\Delta F + T \Delta S $ is a mixture of these two scaling behaviors.

A natural question is of course whether the renormalization
procedure in configuration space that we have developed
 can be somehow 'projected' in real space.
This question is important both numerically and theoretically.
From a numerical point of view, it is clear that the formulation
of RG rules in configuration space has an exponential 'price',
since the number of initial configurations ${\cal N}_0$
grows exponentially with the number of degrees of freedom, 
i.e. grows exponentially with the volume $L^d$ for a system of linear
size $L$ in dimension $d$.
This computational complexity is not surprising,
since the determination of barriers for the dynamics is 
expected to be an NP-complete problem \cite{middleton}.
But it is clear that the numerical study 
of disordered systems via strong disorder
RG rules is then limited to rather small sizes.
From a theoretical point of view, as recalled in the introduction,
it is usual to describe the equilibrium of statistical physics models
by a coarse-graining of some Hamiltonian in real space.
As a consequence of locality of the dynamics, the hierarchy of
valleys that we construct in configuration space is expected to
correspond to some hierarchical organization of real space structures.
It would be very interesting to formulate
a consistent RG procedure directly for these real space structures.
But for the moment it is not clear to us what principle should be used
to construct these correlated real-space structures recursively 
in a consistent way.

\end{document}